\input harvmac
%
%
\message{S-Tables Macro v1.0, ACS, TAMU (RANHELP@VENUS.TAMU.EDU)}
%
%
\newhelp\stablestylehelp{You must choose a style between 0 and 3.}%
\newhelp\stablelinehelp{You 
should not use special hrules when stretching
a table.}%
\newhelp\stablesmultiplehelp{You have tried to place an S-Table  
inside another
S-Table.  I would recommend not going on.}%
%
%
\newdimen\stablesthinline
\stablesthinline=0.4pt
\newdimen\stablesthickline
\stablesthickline=1pt
%
%
\newif\ifstablesborderthin
\stablesborderthinfalse
\newif\ifstablesinternalthin
\stablesinternalthintrue
\newif\ifstablesomit
\newif\ifstablemode
\newif\ifstablesright
\stablesrightfalse
%
%
\newdimen\stablesbaselineskip
\newdimen\stableslineskip
\newdimen\stableslineskiplimit
%
%
\newcount\stablesmode
\newcount\stableslines
\newcount\stablestemp
\stablestemp=3
\newcount\stablescount
\stablescount=0
\newcount\stableslinet
\stableslinet=0
%
%
%
\newcount\stablestyle
\stablestyle=0
%
%
\def\stablesleft{\quad\hfil}%
\def\stablesright{\hfil\quad}%
%
%
\catcode`\|=\active%
%
%
\newcount\stablestrutsize
\newbox\stablestrutbox
\setbox\stablestrutbox=\hbox{\vrule height10pt depth5pt width0pt}
\def\stablestrut{\relax\ifmmode%
                         \copy\stablestrutbox%
                       \else%
                         \unhcopy\stablestrutbox%
                       \fi}%
%
%
\newdimen\stablesborderwidth
\newdimen\stablesinternalwidth
\newdimen\stablesdummy
\newcount\stablesdummyc
\newif\ifstablesin
\stablesinfalse
%
%
\def\begintable{\stablestart%
  \stablemodetrue%
  \stablesadj%
  \halign%
  \stablesdef}%
\def\stablesadj{%
  \ifcase\stablestyle%
    \hbox to \hsize\bgroup\hss\vbox\bgroup%
  \or%
    \hbox to \hsize\bgroup\vbox\bgroup%
  \or%
    \hbox to \hsize\bgroup\hss\vbox\bgroup%
  \or%
    \hbox\bgroup\vbox\bgroup%
  \else%
    \errhelp=\stablestylehelp%
    \errmessage{Invalid style selected, using default}%
    \hbox to \hsize\bgroup\hss\vbox\bgroup%
  \fi}%
\def\stablesend{\egroup%
  \ifcase\stablestyle%
    \hss\egroup%
  \or%
    \hss\egroup%
  \or%
    \egroup%
  \or%
    \egroup%
  \else%
    \hss\egroup%
  \fi}%
\def\stablestart{%
  \ifstablesin%
    \errhelp=\stablesmultiplehelp%
    \errmessage{An S-Table cannot be placed within an S-Table!}%
  \fi
  \global\stablesintrue%
  \global\advance\stablescount by 1%
  \message{<S-Tables Generating Table \number\stablescount}%
  \begingroup%
  \stablestrutsize=\ht\stablestrutbox%
  \advance\stablestrutsize by \dp\stablestrutbox%
  \ifstablesborderthin%
    \stablesborderwidth=\stablesthinline%
  \else%
    \stablesborderwidth=\stablesthickline%
  \fi%
  \ifstablesinternalthin%
    \stablesinternalwidth=\stablesthinline%
  \else%
    \stablesinternalwidth=\stablesthickline%
  \fi%
  \tabskip=0pt%
  \stablesbaselineskip=\baselineskip%
  \stableslineskip=\lineskip%
  \stableslineskiplimit=\lineskiplimit%
  \offinterlineskip%
  \def\borderrule{\vrule width \stablesborderwidth}%
  \def\internalrule{\vrule width \stablesinternalwidth}%
  \def\thinline{\noalign{\hrule height \stablesthinline}}%
  \def\thickline{\noalign{\hrule height \stablesthickline}}%
  \def\trule{\omit\leaders\hrule height \stablesthinline\hfill}%
  \def\ttrule{\omit\leaders\hrule height \stablesthickline\hfill}%
  \def\tttrule##1{\omit\leaders\hrule height ##1\hfill}%
  \def\stablesel{&\omit\global\stablesmode=0%
    \global\advance\stableslines by 1\borderrule\hfil\cr}%
  \def\el{\stablesel&}%
  \def\elt{\stablesel\thinline&}%
  \def\eltt{\stablesel\thickline&}%
  \def\elttt##1{\stablesel\noalign{\hrule height ##1}&}%
  \def\elspec{&\omit\hfil\borderrule\cr\omit\borderrule&%
              \ifstablemode%
              \else%
                \errhelp=\stablelinehelp%
                \errmessage{Special ruling will not display properly}%
              \fi}%
  \def\stmultispan##1{\mscount=##1 \loop\ifnum\mscount>3
\stspan\repeat}%
  \def\stspan{\span\omit \advance\mscount by -1}%
  \def\multicolumn##1{\omit\multiply\stablestemp by ##1%
     \stmultispan{\stablestemp}%
     \advance\stablesmode by ##1%
     \advance\stablesmode by -1%
     \stablestemp=3}%
  \def\multirow##1{\stablesdummyc=##1\parindent=0pt\setbox0\hbox\bgroup%
    \aftergroup\emultirow\let\temp=}
  \def\emultirow{\setbox1\vbox to\stablesdummyc\stablestrutsize%
    {\hsize\wd0\vfil\box0\vfil}%
    \ht1=\ht\stablestrutbox%
    \dp1=\dp\stablestrutbox%
    \box1}%
  
\def\stpar##1{\vtop\bgroup\hsize ##1%
     \baselineskip=\stablesbaselineskip%
     \lineskip=\stableslineskip%
      
\lineskiplimit=\stableslineskiplimit\bgroup\aftergroup\estpar\let\temp=}%
  \def\estpar{\vskip 6pt\egroup}%
  \def\stparrow##1##2{\stablesdummy=##2%
     \setbox0=\vtop to ##1\stablestrutsize\bgroup%
     \hsize\stablesdummy%
     \baselineskip=\stablesbaselineskip%
     \lineskip=\stableslineskip%
     \lineskiplimit=\stableslineskiplimit%
     \bgroup\vfil\aftergroup\estparrow%
     \let\temp=}%
  \def\estparrow{\vfil\egroup%
     \ht0=\ht\stablestrutbox%
     \dp0=\dp\stablestrutbox%
     \wd0=\stablesdummy%
     \box0}%
  \def|{\global\advance\stablesmode by 1&&&}%
  \def\|{\global\advance\stablesmode by 1&\omit\vrule width 0pt%
         \hfil&&}%
  \def\vt{\global\advance\stablesmode by 1&\omit\vrule width  
\stablesthinline%
          \hfil&&}%
  \def\vtt{\global\advance\stablesmode by 1&\omit\vrule width
\stablesthickline%
          \hfil&&}%
  \def\vttt##1{\global\advance\stablesmode by 1&\omit\vrule width ##1%
          \hfil&&}%
  \def\vtr{\global\advance\stablesmode by 1&\omit\hfil\vrule width%
           \stablesthinline&&}%
  \def\vttr{\global\advance\stablesmode by 1&\omit\hfil\vrule width%
            \stablesthickline&&}%
  \def\vtttr##1{\global\advance\stablesmode by 1&\omit\hfil\vrule  
width ##1&&}%
  \stableslines=0%
  \stablesomitfalse}
\def\stablesdef{\bgroup\stablestrut\borderrule##\tabskip=0pt plus 1fil%
  &\stablesleft##\stablesright%
  &##\ifstablesright\hfill\fi\internalrule\ifstablesright\else\hfill\fi%
  \tabskip 0pt&&##\hfil\tabskip=0pt plus 1fil%
  &\stablesleft##\stablesright%
  &##\ifstablesright\hfill\fi\internalrule\ifstablesright\else\hfill\fi%
  \tabskip=0pt\cr%
  \ifstablesborderthin%
    \thinline%
  \else%
    \thickline%
  \fi&%
}%
\def\endtable{\advance\stableslines by 1\advance\stablesmode by 1%
   \message{- Rows: \number\stableslines, Columns:   
\number\stablesmode>}%
   \stablesel%
   \ifstablesborderthin%
     \thinline%
   \else%
     \thickline%
   \fi%
   \egroup\stablesend%
\endgroup%
\global\stablesinfalse}
%

\lref\witdiv{E. Witten, ``String Theory Dynamics in Various
Dimensions,'' Nucl. Phys. {\bf B443} (1995) 85, hepth/9503124.}
\lref\witco{E. Witten, ``Some Comments on String Dynamics,''
hepth/9507121.}
\lref\aspgi{P. Aspinwall and M. Gross, ``Heterotic-Heterotic Duality
and Multiple $K_3$ Fibrations,'' hep-th/9602118.}
\lref\witsi{E. Witten, ``Small Instantons in String Theory,''
Nucl. Phys. {\bf B460} (1996) 541, hepth/9511030.}
\lref\bsvo{M. Bershadsky, V. Sadov, and C. Vafa, ``D Strings on
D Manifolds,'' Nucl. Phys. {\bf B463} (1996) 398, hepth/9510225.}
\lref\stro{A. Strominger, ``Massless Black Holes and Conifolds in
String Theory,'' Nucl. Phys. {\bf B451} (1995) 96, hepth/9504090.}
\lref\dmo{M. Douglas and G. Moore, ``D-branes, Quivers, and
ALE Instantons,'' hepth/9603167.}
\lref\kv{S. Kachru and C. Vafa, ``Exact Results for N=2
Compactifications of Heterotic Strings,'' Nucl. Phys. {\bf B450}
(1995) 69, hepth/9505105.}
\lref\ganor{O. Ganor and A. Hanany, ``Small $E_8$ Instantons and
Tensionless Noncritical Strings,'' hepth/9602120.}
\lref\sw{
 N. Seiberg and E. Witten, ``Comments on String Dynamics in Six
Dimensions,'' hepth/9603003.}  \lref\witphase{E. Witten, ``Phase
Transitions in M Theory and F Theory,'' hepth/9603150.}
\lref\fhsv{S. Ferrara, J. Harvey, A. Strominger, and C. Vafa,
``Second-Quantized Mirror Symmetry,'' Phys. Lett. {\bf B361} (1995)
59, hepth/9505162.}  \lref\mv{D. Morrison and C. Vafa,
``Compactifications of F-theory on Calabi-Yau Threefolds - I,II''
hepth/9602114, hepth/9603161.}  \lref\tate{J. Tate, ``Algorithm for
Determining the Type of a Singular Fiber in an Elliptic Pencil,'' in
{\it Modular Functions of One Variable IV}, Lecture Notes in
Math. vol. 476, Springer-Verlag, Berlin (1975).}
\lref\ibet{G. Aldazabal, A. Font, L. Ibanez, and F. Quevedo, ``Chains
of N=2, D=4 Heterotic/Type II Duals,'' hepth/9510093.}
\lref\canf{P. Candelas and A. Font, ``Duality Between the Webs of
Heterotic and Type II Vacua,'' hepth/9603170.}
\lref\aspg{P. Aspinwall and M. Gross, ``The SO(32) Heterotic String on
a $K_3$ Surface,'' hepth/9605131.}  \lref\vft{C. Vafa, ``Evidence for
F-theory,'' hepth/9602022.}  \lref\kmp{S.~Katz, D.~Morrison,
R.~Plesser, ``Enhanced Gauge Symmetry in Type II String Theory,''
hepth/9601108.}  \lref\ov{H. Ooguri and C. Vafa, ``Two-Dimensional
Black Hole and Singularities of CY Manifolds,'' Nucl. Phys. {\bf B463}
(1996) 55, hepth/9511164.}  \lref\klkam{P. Berglund, S. Katz,
A. Klemm, and P. Mayr, ``New Higgs Transitions Between Dual N=2 String
Models,'' hepth/9605154.}  \lref\DMW{M. Duff, R. Minasian, and
E. Witten, ``Evidence for Heterotic/Heterotic Duality,''
hepth/9601036.}  \lref\HT{C. Hull and P. Townsend, ``Unity of
Superstring Dualities,'' Nucl. Phys. {\bf B438} (1995) 109,
hepth/9410167.}  \lref\GP{E. Gimon and J. Polchinski, ``Consistency
Conditions for Orientifolds and D Manifolds,'' hepth/9601038.}
\lref\schwarz{J. Schwarz, ``Anomaly-Free Supersymmetric Models in Six
Dimensions,'' Phys. Lett. {\bf B371} (1996) 223, hepth/9512053.}
\lref\sagnotti{A. Sagnotti, ``A Note on the Green-Schwarz Mechanism in
Open String Theories,'' Phys. Lett. {\bf B294} (1992) 196,
hepth/9210127.}  \lref\baty{V. Batyrev, ``Dual Polyhedra and Mirror
Symmetry for Calabi-Yau Hypersurfaces in Toric Varieties,''
J. Alg. Geom. {\bf 3} (1994) 493, alg-geom/9310003.}
\lref\ketal{S. Kachru, A. Klemm, W. Lerche, P. Mayr, and C. Vafa,
``Nonperturbative Results on the Point-Particle Limit of N=2 Heterotic
String Compactifications,'' Nucl. Phys. {\bf B459} (1996) 537,
hepth/9508155.}  \lref\klmvw{A. Klemm, W. Lerche, P. Mayr, C. Vafa,
and N. Warner, ``Selfdual Strings and N=2 Supersymmetric Field
Theory,'' hepth/9604034.}  \noblackbox

\Title{\vbox{
\hbox{HUTP-96/A017}
\hbox{IASSNS-HEP-96/49}
\hbox{RU-96-40}
\hbox{\tt hep-th/9605200}}}
{Geometric Singularities and Enhanced Gauge Symmetries}
\vskip 1cm
\centerline{M. Bershadsky$^{a}$,
K. Intriligator$^{b}$, S. Kachru$^{a}$,}
\centerline{D.R. Morrison$^{c,d}$, V. Sadov$^{b}$, and C. Vafa$^{a}$}
\bigskip
\bigskip\centerline{\it $^{a}$Lyman Laboratory of Physics, Harvard
University, Cambridge, MA 02138}
\bigskip\centerline{\it $^{b}$Institute for Advanced Study,
Princeton, NJ 08540}
\bigskip\centerline{\it $^{c}$Dept. of Physics and Astronomy,
Rutgers University, Piscataway, NJ 08855}
\bigskip\centerline{\it $^{d}$Mathematics Department, Duke University,
Durham, NC 27708}
\vskip .3in

Using ``Tate's algorithm,'' we identify loci in the moduli of F-theory
compactifications
corresponding to enhanced gauge symmetry.  We apply this to test the
proposed F-theory/heterotic dualities in six dimensions.  We recover
the perturbative gauge symmetry enhancements of the heterotic side and
the physics of small $SO(32)$ instantons, and discover new mixed
perturbative/non-perturbative gauge symmetry enhancements.  Upon
further toroidal
compactification to 4 dimensions, we derive the chain of
Calabi-Yau threefolds dual to various Coulomb branches of heterotic
strings.

\Date{May 1996}

\newsec{Introduction}

One of the key observations in the recent advances in understanding
non-perturbative aspects of string theory has been the appreciation
that singular geometries of string compactification can be
re-interpreted in terms of solitonic states which become light.  For
example, in type IIA/B string compactification on $K3$ the A-D-E
singularities are associated with massless/tensionless
particles/strings
\refs{\witdiv, \witco, \bsvo}, while
the conifold singularities of Calabi-Yau manifolds are associated with
massless hypermultiplets \refs{\stro, \bsvo}.  A similar story also
turns out to be true for small instantons of $SO(32)$ heterotic
strings \refs{\witsi,\dmo}.

A better understanding of non-perturbative aspects of string theory
clearly requires an extensive knowledge of the physical
reinterpretation of geometric singularities.  It is the aim of this
paper to find a {\it geometry/physics} dictionary for a limited series
of geometric singularities arising in string compactification.  We
will mainly concentrate on compactification to $d=6$ with $N=1$
supersymmetry (or equivalently $d=4$, $N=2$) but the methods have a
wider range of validity.

The starting point is the type IIA/heterotic dualities observed for
$d=4$, $N=2$ \refs{\kv,\fhsv} and, more precisely, their extensions
\mv\ as F-theory/heterotic dualities in $d=6$, $N=1$.  In particular,
we will concentrate on the physical interpretation of the
singularities of the hypermultiplet moduli in F-theory/heterotic
dualities in $d=6$.  Since a hypermultiplet in $d=6$, $N=1$ remains a
hypermultiplet upon further toroidal compactification to $d=4$, our
results hold for that case as well.  Using a method known as ``Tate's
algorithm'' \tate, we explicitly calculate the geometric singularities
of the F-theory compactification which lead to enhanced gauge
symmetries, and are able to classify these singularities.  In
particular, we will find the singularity realization of all classical
gauge groups, including the non-simply laced $B$ and $C$ series and
$F_4$ and $G_2$ gauge groups.  On the heterotic side, some of the
singularities are mapped to partial restoration of {\it perturbative }
gauge symmetries upon partial `un-Higgsing'.  The match between these
descriptions provides further strong evidence for the proposed
dualities.  This also shows how the singularities encode certain
matter representations of the gauge group.

These results have a number of applications.  In particular, we can
uncover certain aspects of non-perturbative gauge symmetry
enhancements on the heterotic side by using the dictionary we develop
for singularities of F-theory compactifications.  This includes
recovering the physical interpretation of small $SO(32)$ instantons
\witsi\ as well as discovering
new mixed perturbative/non-perturbative gauge symmetry enhancements
for certain heterotic compactifications.  In another direction, having
identified the loci of enhanced gauge symmetries, and upon further
compactification to 4 dimensions, we will get new branches
corresponding to the Coulomb branch of these enhanced gauge
symmetries.  In this way, we will be able to {\it derive} the chain of
Calabi-Yau threefolds which correspond to various choices of Coulomb
branches on the heterotic side.  In this way we begin to map out and
in fact derive the web of $d=4, N=2$ heterotic/type II dualities.
These results explain (and extend) some previous results on this topic
\refs{\ibet,\canf}.

The organization of this paper is as follows: In section 2, we review
the F-theory/heterotic duality of \refs{\vft, \mv}.  In section 3, we
present the basic idea for finding the detailed map of the enhanced
gauge symmetry loci.  This is facilitated by the work of Tate, which
we review (and extend).  In section 4, we compare the loci of
perturbative enhanced gauge symmetries on the F-theory and heterotic
sides and find a perfect match.  In section 5, we apply the dictionary
developed to the case of small instantons of $SO(32)$ heterotic
strings.  In section 6, we consider loci of enhanced gauge symmetries
which are mixtures of perturbative and
non-perturbative gauge groups on the heterotic side.  In section 7, we
consider further toroidal compactification to 4 dimensions, where the
F-theory/heterotic duality turns into $N=2$ type IIA/heterotic
duality.  We identify the new Coulomb branches of perturbative
heterotic strings on the type IIA side by deriving the dual chains of
Calabi-Yau threefolds.  Tate's algorithm turns out to be particularly
helpful in enabling us to derive these results.  In section 8, we
present our conclusions.

As this work was being completed, we received a paper with some
overlapping results \aspg.  While the bulk of the work here,
and in particular the section on the F-theory realization of
small $SO(32)$ instantons was completed independently, we have
used (in \S3) some of the ideas of \aspg\ concerning monodromy.

\newsec{Review of F-theory/heterotic dualities}

In this section we review compactifications of F-theory on elliptic
Calabi-Yau manifolds and the corresponding heterotic duals
\refs{\vft,\mv}.  F-theory can be compactified on elliptic Calabi-Yau
manifolds.  Such compactifications can be interpreted as type IIB
vacua where two things happen: 1-The coupling constant $\tau$, which
is to be identified with the complex modulus of elliptic fiber, varies
over space.  2-It is allowed to undergo $SL(2,{\bf Z})$ monodromies,
which are conjectured to be a symmetry of type IIB strings \HT.  The
first fact implies that, at least in the regions where the coupling is
weak ($\tau_2 >>1$) we can use a perturbative string description.
However, the second fact is a marked departure from type IIB
perturbative string vacua.  This is also similar to M-theory
compactifications: In the same way, we can think of M-theory
compactifications as type IIA compactifications where the coupling is
allowed to vary over the space and also make jumps.  It is the latter
fact which makes the geometric M-theory description more powerful and
the type IIA perspective more limited.

An elliptic Calabi-Yau can be described in the Weierstrass form
\eqn\weish{  y^2=x^3+x  f  + g }
which describes the elliptic fibration
(parameterized by $(y,x)$ subject
to the above equation) over the base $B$,
where $f$ and $g$ are functions on the base.

At some divisors $D_i$ the torus (fiber) degenerates. These
divisors are given by the zero loci of the
discriminant
\eqn\disc{\Delta=4 f^3 +27 g^2~.}
Singularities of the manifold are coded in the polynomials
$f $,  $g $ and
determine
the gauge group  and the matter content of the F-theory
compactification.
One can get a singular locus by adjusting various coefficients in
the polynomials $f$ and $g$.
On the heterotic side this process typically corresponds to
``un-Higgsing'' by turning off charged fields
and restoring some gauge symmetry. In this paper, we compare these two
mechanisms and
establish the  precise dictionary between various
singularities of elliptic fibrations
and gauge symmetry
enhancement.

The types of singularities of elliptic fibrations were classified
by Kodaira. His results are
summarized in the table below.  ${\rm ord}(X)$ denotes the order
of the zero of a polynomial $X$ at the discriminant locus.

Table 1: Kodaira Classification of Singularities
\bigskip
\begintable
 ${\rm ord}(f)$ | ${\rm ord}(g)$ | ${\rm ord}(\Delta)$   |
                               fiber type | singularity type \elt
$\geq 0$ | $\geq 0$ | $0$ | smooth |none \elt
$0$ | $0$ | $n$ |  $I_n$  | $A_{n-1}$ \elt
$\geq 1$ | $1$ | $2$| $II$ | none\elt
$1$ | $\geq 2$ | $3$ |  $III$ | $A_1$ \elt
$\geq 2$ | $2$ | $4$ |   $IV$  | $A_2$\elt
$2$ | $\geq 3$ | $n+6$ |  $I_n ^*$ | $D_{n+4}$ \elt
$\geq 2$ | $3$ | $n+6$ |  $I_n ^*$ | $D_{n+4}$ \elt
$\geq 3$ | $4$ | $8$ |  $IV^*$ | $E_6$ \elt
$3$ | $\geq 5$ | $9$ |  $III^*$  | $E_7$\elt
$\geq 4$ | $5$ | $10$ |  $II^*$  | $E_8$
\endtable
\bigskip
When  ${\rm ord}(f)\geq 4$ and ${\rm ord}(g)\geq 6$,
the singularity of the manifold
is so bad
that it generically destroys the triviality of the canonical bundle.

If the base $B={\bf P}^1$ then $f$ and $g$ are functions of one
variable only, say $z_1$.  This corresponds to compactification of
F-theory on $K3$, which has been conjectured to be dual to
heterotic compactification on $T^2$ \vft.  The duality of this
F-theory compactification with type I has been recently discussed in 
\ref\sen{A. Sen, ``F Theory and Orientifolds,'' hep-th/9605150.}.  
In F-theory on K3, the singularity type of Kodaira
exactly matches the conjectured enhanced gauge symmetry of the
eight-dimensional theory.  This, in fact, can be verified in the
F-theory language for the $A_n$ case by the realization that it
corresponds to $n+1$ parallel 7-branes \vft. 
Upon further compactification on $T^2$ and using
the equivalence with type IIA on $K3$, this realizes the mechanism of
gauge symmetry enhancement for singular $K3$s suggested in
\refs{\bsvo,
\ov}.  For the other types of singularities,
string-string duality in six dimensions requires that we identify
the Weierstrass A-D-E classification with the corresponding gauge
symmetry enhancement \witdiv.  

The meaning of the singularity type when the base is more than one
dimensional is a priori less clear.  We can still continue to use
Kodaira's terminology of A-D-E for the singularity type even when the
dimension of $B$ is bigger than one, but the physical implication of
the singularity turns out to be more intricate.  The main aim of this
paper is to address this issue.  We will mainly concentrate on the case
$B=F_n$, where $F_n$ is the rational ruled surface, but our methods are
more general.

F-theory on the elliptic CY 3-fold over $F_n$ is conjectured to be  
dual to
the $E_8\times E_8$ heterotic string on $K3$ with instanton
numbers $(12+n,12-n)$ in the two $E_8$s \mv .
We can better map the moduli of this
duality if we can identify loci on each side which correspond to
enhanced gauge symmetries.  It turns out to be relatively easy to
identify the loci corresponding to enhanced gauge symmetries on the
F-theory side, for any given group, as we will explain below.
However, it turns out to require somewhat more work to identify the
matter representations involved.  To do that rigorously one can
consider further compactification to 4 dimensions, upon which F-theory
becomes equivalent to type IIA on the CY 3-fold.  One can then study
the wrapping of the D-branes in the type IIA theory to see the  
gauge group
\refs{\bsvo,\kmp} and check the matter
representation as well.  This approach has been considered for some
examples in the recent work \klkam.  Instead of this direct approach,
we use the duality with heterotic strings to develop the matter
multiplet dictionary on the F-theory side.

As mentioned above we will mainly concentrate on
 the Hirzebruch surface $F_n$ as a base for the Calabi-Yau
threefold.
The Hirzebruch surface is a ${\rm \bf P}^1$ fibration over ${\rm
\bf P}^1$
characterized by one integer $n$.  We choose our convention so that
$z_1$ is the coordinate along ${\rm \bf P}^1$ fiber, while $z_2$ is
the coordinate along the base.  To describe the CY 3-fold over  
$F_n$ it is
most convenient to expand the functions $f(z_1,z_2)$ and $g(z_1,z_2)$
of \weish\ in powers of $z_1$:
\eqn\expan{\eqalign{xf(z_1,z_2)=& x\sum _{i=0}^I z_1 ^i
f_{8+n(4-i)}(z_2)  \cr
   g(z_1,z_2)=&\sum_{j=0}^J z_1 ^j g_{12+n(6-j)}(z_2)~,}}
where the subscripts on $f$ and $g$ denote the degree of the
polynomial in $z_2$ and
where $I\leq 8$ is the largest value with $8+n(4-I)\geq 0$ and $J\leq
12$ is the largest value with $12+n(6-J)\geq 0$.  The correlation
between the degree of the polynomials and the power of $z_1$
follows from the fact that $(z_1,z_2)$
parameterize $F_n$.

It has been argued in the second reference in \mv\ that the ``middle
polynomials'', i.e. the coefficients of $xz_1^4$ and $z_1^6$
(corresponding to $i=4$ and $j=6$) in \expan, correspond to the moduli
of the $K3$ on which the heterotic theory is
compactified. Furthermore, it was argued that polynomials of lower
degree in $z$ (i.e., $xz_1^i$ for $i<4$ and $z_1^j$ for $j<6$) control
the moduli of one $E_8$, based near $z_1=0$, with instanton number
$12+n$.  Polynomials of higher degree in $z$ ($i>4$ and $j>6$) control
the other $E_8$, based near $z_1=\infty$, with instanton number
$12-n$.  The zeroes of $g_{12+n}$ and $g_{12-n}$ (the coefficients of
$z_{1}^{5}$ and $z_{1}^{7}$ in \expan) were conjectured, when all
lower/higher terms are set to zero, to correspond to $12+n$ small
instantons in the $z_{1}=0$ $E_8$ and $12-n$ small instantons in the
$z_{1}=\infty$ $E_8$.

We wish to map the perturbative enhanced gauge symmetries of the
heterotic side onto the F-theory moduli.  On the F-theory side, a
perturbative enhanced gauge symmetry of the heterotic theory should
already be part of the gauge symmetry in 8 dimensions.  In other
words, if we consider the size of the second $P^1$ to be big then, as
a function of $z_1$, we should get a singularity on the F-theory side,
reflecting the existence of gauge symmetry already in 8-dimensions.
We thus restrict our attention to the singularities at $z_1=const$.
We will be mainly focusing on $E_8\times E_8$ heterotic strings.  In
this case, the gauge symmetries are localized on the F-theory side on
two points, $z_1=0,\infty$.  With no loss of generality, we will focus
on the gauge symmetries coming from the $E_8$ at $z_1=0$ with
instanton number $12+n$.  Singularities at $z_2=const$ (or more
general singular loci that cannot be represented as $z_1=const$)
correspond to {\it non-perturbative} gauge symmetry enhancement, as
they are localized on a point in the $z_2$ space which is `visible' to
both the heterotic strings and F-theory.

 Before getting to a more detailed match of the moduli of the $E_8$
bundle, let us see how the total count of the dimension of the moduli
space of $E_8$ bundles works. The dimension of the hypermultiplet
moduli space for $E_8$ with $12+n$ instantons is $30n+112$.  Each F
theory complex structure modulus leads to one hypermultiplet.
Counting the terms in \expan\ with $i<4$ and $j<6$, we find $31n+114$,
but $n+2$ of these are associated with reparameterizations
$z_1\rightarrow az_1+P_n(z_2)$, yielding the correct number of
hypermultiplets.  Requiring the unbroken part of the gauge group to be
at $z_1=0$ fixes the reparameterizations corresponding to shifts by
$P_n(z_2)$.  We then only have to subtract one (corresponding to
rescaling $z_1$) from the total count of the parameters in the
polynomials to obtain the dimension of the moduli of the bundles.

\newsec{Basic Idea and Tate's algorithm}

Before getting to specific cases, we wish to give the general picture:
We start with an A-D-E type singularity at $z_1=0$.  We then consider a
${\rm \bf P}^1$ fibration of this, given by the coordinate $z_2$.
Then, depending on what this fibration is, we will in general `break'
some of the symmetries that we start with.  We thus expect that, if we
start with an A-D-E singularity, we end up with a gauge symmetry which
is a subgroup of that of the singularity type.  The `breaking' is
actually very restrictive: As we go around a point on the $z_2$ plane
where the fiber degenerates, we come back to the same singularity in
the fiber up to a monodromy action on the singularity.  If the
monodromy action is given by a Weyl group element, i.e. if it is an
{\it inner} automorphism, we can undo it by a gauge transformation in
the fiber.  However, if the monodromy action is not given by a Weyl
group element, i.e. if it is an {\it outer} automorphism, it cannot be
undone by a gauge transformation.  As we shrink the cycle to zero
size, we end up orbifolding the gauge group by an outer automorphism,
which thus means that we have reduced gauge symmetry \aspg.

It is well known that the actions of outer automorphisms are
realizable as automorphisms of the A-D-E Dynkin diagrams.  The main
ones we will encounter in this paper are given by
\eqn\outera{\eqalign{A_{2n-1}&\rightarrow C_n\cr
D_n&\rightarrow B_{n-1}\cr
E_6&\rightarrow F_4\cr
D_4&\rightarrow G_2,}}
where all but the last one are involutions and the last one is the
triality automorphism of $D_4$.  We shall find below that all of these
cases are essentially realized.  In fact, the best way to phrase them
in the singularity language is as follows: Suppose we resolve the
singularity of the Calabi-Yau manifold.  Then we get the A-D-E Dynkin
diagram corresponding to the intersection diagram of vanishing
2-cycles.  As we fiber this space over another parameter $z_2$ and go
over non-trivial cycles, in general the vanishing cycles come back to
themselves but may undergo monodromy.  If these vanishing cycles are
exchanged according to an outer automorphism of the Dynkin diagram,
the actual gauge group will be smaller as indicated above.  In fact,
the distinction between the types of A-D-E singularities occurring
in elliptic fibrations
according to whether the vanishing cycles mix, which we call {\it
non-split}, or do not mix, which we call {\it split} is implicit in
the work of
Tate \tate.  Generically the A-D-E fibrations will undergo outer
automorphisms, i.e. we have the non-split case, which means that for
a generic fibration of the A-D-E type we will get the right hand side
above as the gauge symmetry.  However, if we put some extra conditions
on the fibration, in accord with Tate's algorithm, we can avoid this
breaking and get back the original group.  This will be seen in the
examples below.

It is important to keep in mind that a given A-D-E singularity may
$\it not$ correspond to an enhanced gauge symmetry in the F-theory.
One example is $E_8$ fiber singularities, which seem to correspond to
small instantons but not $E_8$ gauge symmetry \mv.  We will find that
all the other cases noted above, on the other hand, do occur as
expected.  There are other outer automorphisms which do not occur,
however.  For example the non-split $A_{2k}$ singularity, which would
be expected to give $Sp(k)$ gauge symmetry\foot{There is a subtle
argument for this having to do with the fact that the outer
automorphism acting on the Lie algebra has order 4
\ref\jle{J. Lepowski, private communication.}.}, does not seem to give
rise to conventional gauge symmetries.  Presumably this is similar to
the situation which arises when $E_8$ instantons shrink to zero size
\refs{\ganor,\sw,\witphase}.

We will now review (and extend) Tate's algorithm.

\subsec{Tate's algorithm}

Tate's algorithm \tate\ gives a procedure for computing the Kodaira  
type of a
singular fiber in an elliptic fibration (as well as the
``split/nonsplit'' distinction mentioned in the previous subsection),
at a generic point along any divisor $\Sigma$ in the base $B$ of the
fibration.  The basic idea is a refinement of the method already  
mentioned,
and proceeds by
studying the order of vanishing of the coefficients of the defining
equation along $\Sigma$.

In order to carry this program out in an efficient manner, it is
necessary to begin with a more general form of the Weierstrass
equation.  We take the form to be
\eqn\egenW{y^2+a_1xy+a_3y=x^3+a_2x^2+a_4x+a_6,}
where the $a_i$'s are locally defined polynomial functions on the base
(or more generally, sections of line bundles).  The traditional
Weierstrass form \weish\ can be obtained from \egenW\ by completing
the square in $y$ and then completing the cube in $x$.

The algorithm makes reference to several quantities derived from the
coefficients $a_i$, defined by
\eqn\bees{\eqalign{
 b_2&=a_1^2+4a_2\cr
 b_4&=a_1a_3 +2a_4\cr
 b_6&=a_3^2+4a_6 \cr
 b_8&=b_2a_6-a_1a_3a_4+a_2a_3^2-a_4^2 \cr
\Delta&=-b_2^2b_8-8b_4^3-27b_6^2+9b_2b_4b_6. \cr
}}
The discriminant $\Delta$ is the same one used earlier, up to a
numerical factor.  For later reference, we also record the results
of completing the square and the cube:
\eqn\fg{\eqalign{
f&=-{1\over48}(b_2^2-24b_4)\cr
g&=-{1\over864}(-b_2^3+36b_2b_4-216b_6).\cr
}}

In carrying out the algorithm, we let $\{\sigma=0\}$ be a local
defining equation for $\Sigma$, and we examine various divisibility
conditions of the form `$\sigma^k$ divides $a_j$'.  When such
a condition holds, we define
$a_{j,k}=a_j/\sigma^k$, a notation which will be used throughout this
discussion.  We similarly define $b_{j,k}=b_j/\sigma^k$ whenever it
makes sense.

The algorithm now proceeds roughly as follows: locate the  
singularity in the
fibers over $\Sigma$, make a change of coordinates in $(x,y)$ to put
the singularity in a convenient location, blow up the singularity, and
then repeat.  At each stage in this process, after the change of
coordinates has been made, the coefficients in the equation will be
divisible by certain powers of the local defining equation $\sigma$.
This can be used to characterize which branch of the algorithm should
be used.

For example, in the first step of the algorithm we ask whether the
fibers over $\Sigma$ are actually singular, i.e., whether $\sigma$
divides $\Delta$.  If so, we change
coordinates to put the singularity at $(x,y)=(0,0)$.  When the
singularity is located there, we have $\sigma$ dividing $a_3$, $a_4$
and $a_6$, and generically the Kodaira fiber is of type I$_1$ (no
enhanced gauge symmetry).
The discriminant mod $\sigma^2$ can be calculated as
$$ \Delta \equiv - \sigma \, b_2^3 \, a_{6,1} \pmod{\sigma^2}.$$

Any worsening of the singularities
is thus indicated by either $b_2\equiv0$ mod
$\sigma$, or by $a_{6,1}\equiv0$ mod $\sigma$.  The first case leads to
the branch in the algorithm giving fibers of Kodaira types II, III,
etc., while the second case leads to I$_n$ type fibers.

To see how the split/non-split distinction arises in this algorithm,
we need to proceed a bit further
along the I$_n$ branch.  So suppose that
in addition to the previous conditions,
 $\sigma^2$ divides $a_6$.  Then
the singular point of the fiber is also a singular point
of the total space of the fibration,
 and
we should blow up the origin in $(x,y,\sigma)$ space to begin
resolving the singularity.
The leading order terms in the equation are
\eqn\elead{ y^2 + a_1 xy + a_{3,1} \sigma y = a_2 x^2 + a_{4,1}  
\sigma x +
a_{6,2}
\sigma^2 ;}
if this quadratic is nonsingular, then blowing up the origin in
$(x,y,\sigma)$ resolves the singularity which was an $A_1$.  We have
in this case
Kodaira type I$_2$, and gauge group $SU(2)$.  The discriminant mod
$\sigma^3$ can be calculated as
$$ \Delta = - \sigma^2 \, b_2^2 \, b_{8,2} \pmod{\sigma^3},$$
where $b_{8,2}$ coincides with
 the discriminant of the quadratic equation \elead\ (a
cubic in the coefficients).  To get worse singularities, either
$b_2$ or $b_{8,2}$  should
vanish mod $\sigma$.

The split/non-split distinction arises at the next branch in the
algorithm.  If in addition to the previous assumptions, we assume
$b_{8,2}$ is divisible by $\sigma$, but $b_2$ is not,
 then \elead\ is a quadratic
equation of rank precisely two, and by a change of coordinates we
may assume that \elead\ involves $x$ and $y$ alone, i.e., that the
entire equation takes the form
$$ y^2 + a_1 xy + a_{3,2} \sigma^2 y = a_2 x^2 + a_{4,2} \sigma^2 x +
a_{6,2}
\sigma^2 .$$
The exceptional divisor of the blowup map is defined by
\eqn\eleadd{
y^2 + a_1 xy - a_2 x^2 \equiv0 \pmod\sigma}
which consists of two lines for each specific numerical value of the
coefficients $a_{j,k}$.  However, since those coefficients will
actually depend on the other parameters in the base, it may not be
possible to individually define those two lines globally along
$\Sigma$.  In order for it to be possible, we need a factorization
$$y^2 + a_1 xy - a_2 x^2\equiv
(y-rx)(y-sx) \pmod\sigma$$
for some functions $r$ and $s$ on the base.  If this factorization
exists, then after a change of coordinates in $y$ we may assume that
$r\equiv0$ mod $\sigma$, i.e., that $\sigma$ divides $a_2$.  This case
gives Kodaira fiber I$_3^s$ and gauge group $SU(3)$.
(We use the superscripts $ns$ and $s$ on Kodaira fibers to distinguish
between the non-split and split cases.)  If the
factorization does not exist, then we are in the `non-split' situation
with a singularity of type $A_2$; this is the case in which the gauge
symmetry is unconventional.  (The Kodaira fiber is I$_3^{ns}$.)
Further down this branch of the algorithm
one encounters the $Sp(k)$ gauge groups.

Rather than continuing through the algorithm step by step, we have
summarized the conditions it entails in table 2.
\vfill
\eject

Table 2: Tate's Algorithm \bigskip \begintable type | group | $ a_1$ |
$a_2$ | $a_3$ |$ a_4 $| $ a_6$ |$\Delta$ \elt $I_0 $ | --- |$ 0 $ |$ 0
$ |$ 0 $ |$ 0 $ |$ 0$ |$0$ \elt $I_1 $ | --- |$0 $ |$ 0 $ |$ 1 $ |$ 1
$ |$ 1 $ |$1$ \elt $I_2 $ |$SU(2)$ |$ 0 $ |$ 0 $ |$ 1 $ |$ 1 $ |$2$ |$
2 $ \elt $I_{3}^{ns} $ | unconven. |$0$ |$0$ |$2$ |$2$ |$3$ |$3$ \elt
$I_{3}^{s}$ |unconven. |$0$ |$1$ |$1$ |$2$ |$3$ |$3$ \elt
$I_{2k}^{ns}$ |$ Sp(k)$ |$0$ |$0$ |$k$ |$k$ |$2k$ |$2k$ \elt
$I_{2k}^{s}$ |$SU(2k)$ |$0$ |$1$ |$k$ |$k$ |$2k$ |$2k$ \elt
$I_{2k+1}^{ns}$ |unconven. | $0$ |$0$ |$k+1$ |$k+1$ |$2k+1$ |$2k+1$
\elt $I_{2k+1}^s$ |$SU(2k+1)$ |$0$ |$1$ |$k$ |$k+1$ |$2k+1$ |$2k+1$
\elt $II$ | --- |$1$ |$1$ |$1$ |$1$ |$1$ |$2$ \elt $III$ |$SU(2)$ |$1$
|$1$ |$1$ |$1$ |$2$ |$3$ \elt $IV^{ns} $ |unconven. |$1$ |$1$ |$1$
|$2$ |$2$ |$4$ \elt $IV^{s}$ |$SU(3)$ |$1$ |$1$ |$1$ |$2$ |$3$ |$4$
\elt $I_0^{*\,ns} $ |$G_2$ |$1$ |$1$ |$2$ |$2$ |$3$ |$6$ \elt
$I_0^{*\,ss}$ |$SO(7)$ |$1$ |$1$ |$2$ |$2$ |$4$ |$6$ \elt $I_0^{*\,s}
$ |$SO(8)^*$ |$1$ |$1$ |$2$ |$2$ |$4$ | $6$ \elt $I_{1}^{*\,ns}$
|$SO(9)$ |$1$ |$1$ |$2$ |$3$ |$4$ |$7$ \elt $I_{1}^{*\,s}$ |$SO(10) $
|$1$ |$1$ |$2$ |$3$ |$5$ |$7$ \elt $I_{2}^{*\,ns}$ |$SO(11)$ |$1$ |$1$
|$3$ |$3$ |$5$ |$8$ \elt $I_{2}^{*\,s}$ |$SO(12)^*$ |$1$ |$1$ |$3$
|$3$ |$5$|$8$\elt 
$I_{2k-3}^{*\,ns}$ |$SO(4k+1)$ |$1$ |$1$ |$k$ |$k+1$
|$2k$ |$2k+3$ \elt $I_{2k-3}^{*\,s}$ |$SO(4k+2)$ |$1$ |$1$ |$k$ |$k+1$
|$2k+1$ |$2k+3$ \elt $I_{2k-2}^{*\,ns}$ |$SO(4k+3)$ |$1$ |$1$ |$k+1$
|$k+1$ |$2k+1$ |$2k+4$ \elt $I_{2k-2}^{*\,s}$ |$SO(4k+4)^*$ |$1$ |$1$
|$k+1$ |$k+1$ |$2k+1$ 
|$2k+4$ \elt $IV^{*\,ns}$ |$F_4 $ |$1$ |$2$ |$2$ |$3$ |$4$
|$8$\elt $IV^{*\,s} $ |$E_6$ |$1$ |$2$ |$2$ |$3$ |$5$ | $8$\elt
$III^{*} $ |$E_7$ |$1$ |$2$ |$3$ |$3$ |$5$ | $9$\elt $II^{*} $
|$E_8\,$ |$1$ |$2$ |$3$ |$4$ |$5$ | $10$ \elt
 non-min | --- |$ 1$ |$2$ |$3$ |$4$ |$6$ |$12$
\endtable

\vfill
\eject

Table 2, con.: Tate's Algorithm, Discriminant and Next Branches
\bigskip
\begintable
type | $\widehat\Delta$ mod $\sigma$ |  next branches \elt
I$_0$| $\Delta$ | I$_1$ \elt
I$_1$ | $-b_2^3 \, a_{6,1}$ |
   I$_2$, II \elt
I$_2$ | $-b_2^2 \, b_{8,2} $ |
   I$_3^{ns}$, III \elt
I$_{3}^{ns}$ |
$-b_2^3\,a_{6,3}$ | I$_{4}^{ns}$, (I$_{3}^{s}$) \elt
I$_{3}^{s}$ | $-a_1^6\,a_{6,3}$
| I$_{4}^{s}$, IV$^{s}$ \elt
I$_{2k}^{ns}$|
$-b_2^2\,b_{8,2k}$ | I$_{2k+1}^{ns}$, I$_{2k}^{s}$ \elt
I$_{2k}^{s}$ | $-a_1^4\,b_{8,2k}$
| I$_{2k+1}^{s}$, (I$_{2k-4}^{*\,s}$) \elt
I$_{2k+1}^{ns}$|
$-b_2^3\,a_{6,2k+1}$ | I$_{2k}^{ns}$, (I$_{2k+1}^{s}$) \elt
I$_{2k+1}^{s}$| $-a_1^6\,a_{6,2k+1}$
| I$_{2k+2}^{s}$, I$_{2k-3}^{*\,s}$ \elt
II|$ -432\,a_{6,1}^2$ | III \elt
III | $ -64\,a_{4,1}^3$ | IV$^{ns}$ \elt
IV$^{ns}$| $ -27\,b_{6,2}^2$ |
IV$^{s}$ \elt
IV$^{s}$ | $ -27\,a_{3,2}^4$ |
I$_0^{*\,ns}$ \elt
I$_0^{*\,ns}$| $\Delta/\sigma^6$ | I$_0^{*\,ss}$  \elt
I$_0^{*\,ss}$ | $16\,a_{4,2}^2(a_{2,1}^2{-}4a_{4,2})$ |
I$_0^{*\,s}$, I$_1^{*\,ns}$   \elt
I$_0^{*\,s}$
|$16\,a_{4,2}^2(a_{2,1}^2{-}4a_{4,2})$
| (I$_1^{*\,ns}$) \elt
I$_{1}^{*\,ns}$ |$-16\,a_{2,1}^3\,b_{6,4}$
| I$_{1}^{*\,s}$, IV$^{*\,ns}$  \elt
I$_{1}^{*\,s}$|$-16\,a_{2,1}^3\,a_{3,2}^2$
| I$_{2}^{*\,ns}$, IV$^{*\,s}$  \elt
I$_{2}^{*\,ns}$ |
$16\,a_{2,1}^2\,(a_{4,3}^2{-}4a_{2,1}a_{6,5})$
| I$_{2}^{*\,s}$, III$^*$  \elt
I$_{2}^{*\,s}$ |
$16\,a_{2,1}^2\,(a_{4,3}^2{-}4a_{2,1}a_{6,5})$
| I$_{2k+1}^{*\,ns}$ , (III$^*$) \elt
I$_{2k-3}^{*\,ns}$|
$-16\,a_{2,1}^3\,b_{6,2k}$
| I$_{2k-3}^{*\,s}$, non-min  \elt
I$_{2k-3}^{*\,s}$ |
$-16\,a_{2,1}^3\,a_{3,k}^2$
| I$_{2k-2}^{*\,ns}$, non-min  \elt
I$_{2k-2}^{*\,ns}$|
$16\,a_{2,1}^2\,(a_{4,k+1}^2{-}4a_{2,1}a_{6,2k+1})$
| I$_{2k-2}^{*\,s}$, non-min  \elt
I$_{2k-2}^{*\,s}$ |
$16\,a_{2,1}^2\,(a_{4,k+1}^2{-}4a_{2,1}a_{6,2k+1})$
| I$_{2k-1}^{*\,ns}$, non-min  \elt
IV$^{*\,ns}$|
$-27\,b_{6,4}^2$ | IV$^{*\,s}$  \elt
IV$^{*\,s}$ |
$-27\,a_{3,2}^4$ | III$^*$  \elt
III$^{*}$|
$-64\,a_{4,3}^3$ | II$^*$  \elt
II$^{*}$ |
$-432\,a_{6,5}^2$ |  non-min \elt
non-min|$\Delta/\sigma^{12}$ |---
\endtable

 This table
is to be interpreted as follows:  If upon change of coordinates in
$(x,y)$, the coefficients in the equation are divisible by
the given powers of $\sigma$,
but the coefficients are otherwise generic, then the Kodaira fiber has
the stated type\foot{The final entry in the table labeled  
``non-min'' refers to
elliptic
fibrations whose singularities are sufficiently bad as to destroy
the ``trivial canonical bundle'' property of the total space
of the fibration---in the mathematics literature, these are called
non-minimal Weierstrass fibrations.}
and we predict the enhanced gauge symmetry stated
in the second column.  The $*$ next to the
 $SO(8), SO(12)$ and $SO(4k+4)$ cases signifies that in addition to the
conditions specified in table 2 a factorization condition
must be satisfied to obtain the enhanced gauge symmetry:
For $SO(8)$ the polynomial
$$X^2 + a_{2,1}X + a_{4,2}$$
should factor modulo $\sigma$, while for the other $SO(4k+4)$ cases  
(including
$SO(12)$)
the polynomial
$$a_{2,1}X^{2} + a_{4,k+1}X + a_{6,2k+1}$$
should factor modulo $\sigma$.
The order of vanishing of the
discriminant is also shown.  Worse singularities occur if
$\widehat\Delta=\Delta/\sigma^{ord(\Delta)}$ satisfies
$\widehat\Delta\equiv0$ mod $\sigma$ or if a factorization
condition is satisfied, as in the $SO(4k+4)$ case
noted above.  The second portion of the table exhibits
$\widehat\Delta$, and indicates which branch of the algorithm is to be
followed when worse singularities occur.

We will see in section 7 how the conditions stated in the table are
used to describe the resolution of singularities by blowing up.  Let
us comment here a bit further about the ``factorization conditions''
which lead to the distinctions between non-split and split cases.
These distinctions, as well as the gauge groups in the second column,
are not explicitly present in Tate's paper, although the rest of the
algorithm is.

1) In the case
of I$_n$, the polynomial whose factorization is at issue takes the
form
$$Y^2 + a_1XY -a_2X^2$$
and (as we already saw in the case $n=3$) whenever it factors
(mod $\sigma$) we can
make a change of coordinates to make one of the factors be $Y$, i.e.
to make $\sigma$ divide $a_2$.

2) In the cases of types IV, IV$^*$, and I$_{2k-3}^*$, the polynomial
whose factorization is at issue takes the form
$$Y^2 + a_{3,k}Y -a_{6,2k}$$
and similarly in this case, a change of coordinates allows us to take
one of the factors (mod $\sigma$) to be $Y$.

3) In the case of type I$_0^*$, the polynomial which needs to be
factored is
$$X^3 + a_{2,1}X^2 + a_{4,2}X + a_{6,3}.$$
There are three cases: no factorization (non-split), full
factorization as a product of three linear factors (split), or
factorization into a linear factor and a quadratic factor.  We refer
to this last case as ``semi-split'' and denote the corresponding
Kodaira fiber as I$_0^{*\,ss}$.  The Dynkin diagram for the
singularity is $D_4$; the non-split case corresponds to the quotient
by $S_3$, yielding the $G_2$ diagram, whereas the semi-split case
corresponds to the quotient by $Z_2$, yielding the $B_3$ diagram.
When there is a linear factor, a coordinate change can be used to make
this factor be $X$, but in the split case only one of the three
factors can be so shifted and we must formulate the condition in terms
of whether a polynomial can be factored mod $\sigma$.

4) Finally, in the case of type I$_{2k-2}$, the polynomial we must
consider is
$$ a_{2,1}X^2 + a_{4,k+1}X + a_{6,2k+1}.$$
Even when this factors, we cannot make a compensating shift of
coordinates.  For one thing, the factorization involves a
factorization of $a_{2,1}$ mod $\sigma$, which is itself a function
(or a section of a line bundle) on
the curve $\Sigma$.  In fact, the flexibility to choose this
factorization differently allows for a number of different types of
these singularities, as we shall see in the examples.

\newsec{Higgs Branches}

We now begin our detailed comparison of the F-theory and heterotic
loci of enhanced gauge symmetries.  We start with the
enhanced gauge symmetries which are realized perturbatively on the
heterotic side.  We focus on the $E_8$ with $12+n$ instantons and
obtain a dictionary of the correspondence between F-theory geometric
singularities and gauge symmetry.  The dictionary, of course, also
directly applies to the $E_8$ with $12-n$ instantons as well as to
situations with additional tensor multiplets.

$E_8$ with $12+n$ instantons has a $30n+112$ dimensional space of
hypermultiplet moduli associated with the gauge bundle.  $E_8$ is
generically broken on this moduli space, with subgroups un-Higgsed on
various subspaces.  An enhanced gauge symmetry $G\subset E_8$
corresponds to restricting the $12+n$ instantons to sit in $H\subset
E_8$ which is the commutant of $G$: $G\times H\subset E_8$ is a
maximal subgroup.  The dimension of the subspace of enhanced $G$ gauge
symmetry, corresponding to the number of $G$ neutral hypermultiplet
moduli, is given by\foot{This and the next equation apply for $H$
simple.  When, as will be the case in some of the examples which
follow, $H$ is a product of simple factors, these formulae are
modified in an obvious fashion.}
\eqn\ine{{\rm dim}~{\cal M}(G)=c_2(H)(12+n)- {\rm dim}~ H}
where $c_2(H)$ is the dual Coxeter number of $H$; this is the
dimension in units of hypermultiplets (one quarter of the real
dimension) for $12+n$ instantons embedded in $H$.

In addition to the above neutral hypermultiplets, on the subspace with
enhanced $G$ gauge symmetry there are massless matter hypermultiplets
transforming in representations $R_i$ of $G$.  The number $N_i$ of
matter fields in representation $R_i$ of $G$ is given by an index
theorem to be
\eqn\nmatteri{N_i=(12+n){\rm index}(S_i)-{\rm dim}(S_i),}
where $S_i$ is the representation of $H$ entering in the decomposition
$adj\ (E_8)=\sum _i (R_i,S_i)$ and the last term is a gravitational
contribution which takes into account the compactification on $K3$.
``Relaxing'' the instantons to instead lie in $H' \supset H$
corresponds to breaking $G$ to commutant $G'\subset G$ by the Higgs
mechanism, giving an expectation value to some of the $G$
charged matter.

The entire moduli space has an intricate structure, with a variety of
subspaces with enhanced gauge symmetry corresponding to the different
possible Higgs mechanisms.  We will organize our discussion by
following two different chains of the Higgs mechanism: One starting
with unbroken $E_7$, corresponding to instantons in $H=SU(2)$, and one
starting from unbroken $SO(12)$, corresponding to instantons in
$H=SU(2)\times SU(2)$.  Upon Higgsing, as will be discussed, these two
connect at various places.  One could consider, more generally, Higgs
chains starting from instantons in $H\subset E_8$ consisting of more
$SU(2)$ factors.  The number of possible factors depends on $n$
because, as follows from \nmatteri, each $SU(2)$ factor must have at
least 4 instantons.

The intricate structure of various enhanced gauge symmetries will be
exactly matched by the geometric singularities of F-theory, with the
dimensions of enhanced gauge symmetry subspaces {\it perfectly}
matching the dimensions of the moduli spaces of compactifications with
various singularities in F-theory.  Moreover, we will use this
dictionary to deduce on the F-theory side which matter representations
are present and how they are encoded in the singularity.  It is
natural to expect that the matter comes from intersecting loci of
singularities, i.e. at points on the $z_2$ plane where the fiber has a
worse singularity, which are reflected as extra zeroes of the
discriminant.  For example, such is the case in the context of
singularities realizable as intersecting D-branes \bsvo .  We find
that this is indeed the case, at least for simply laced gauge groups.
In the case where the unbroken gauge group is non-simply laced, it
seems difficult to `localize' the matter at the intersection points of
the singularity.

\vfill
\eject

The portion of the web of vacua that we have explored is summarized in
table 3 and diagram 1 below.  The details are provided in the rest  
of this
section.
The superscripts $s$, $ns$, and $ss$ in table 3 stand
for split, nonsplit, and semi-split, in the terminology of 
\S3.

Table  3: Various Higgs Branches
\bigskip
\begintable
 Type | Group | Matter content | Dim($\cal M$)\elt $E_7$ | $E_7$ |
 $({n \over 2}+4){\rm \bf 56}$ | $2n+21$ \elt $E_6^{s}$ | $E_6$
 |$(n+6){\rm \bf 27}$ | $3n+28$ \elt $E_6^{ns}$ | $F_4$ | $(n+5){\rm
 \bf 26}$| $4n+34$ \elt $D_5^{s}$ | $SO(10)$ | $(n+4){\rm \bf
 16}+(n+6){\rm \bf 10}$ | $4n+33$ \elt $D_5^{ns}$ | $SO(9)$ |
 $(n+5){\rm \bf 9}+(n+4){\rm \bf 16}$ | $5n+39$ \elt $D_4^{s}$ |
 $SO(8)$ | $(n+4)({\rm \bf 8}_c+{\rm \bf 8}_s+{\rm \bf 8}_v)$ |
 $6n+44$
\elt
 $D_4^{ss}$ | $SO(7)$ | $(n+3){\rm \bf 7}+(2n+8){\rm \bf 8}$ | $7n+48 $
 \elt $D_4^{ns}$ | $G_2$ | $(3n+10){\rm \bf 7}$ | $9n+56$ \elt  
$A_3^{s}$ |
 $SU(4)$ | $(n+2){\rm \bf 6}+(4n+16){\rm \bf 4}$ | $8n+51 $ \elt  
$A_3^{ns}$
 | $SO(5)$ | $(n+1){\rm \bf 5}+(4n+16){\rm \bf 4}$ | $9n+53 $ \elt
$A_1\times A_1$ | $SO(4)$ | $n {\rm \bf (2,2)} + (4n+16) [{\rm \bf  
(1,2) +
(2,1)}]$ | $10n + 54$ \elt
 $A_2^{s}$ | $SU(3)$ | $(6n+18){\rm \bf 3}$ | $12n+66 $ \elt $A_1$ |
 $SU(2)$ | $(6n+16){\rm \bf 2}$ | $18n+83 $\elt
$A_1$ | $SU(2)_2$ | $(8n+32){\rm \bf 2} + (n-1) {\rm \bf 3}$ | $11n  
+ 54$ \elt
 $D_6^{s}$ | $SO(12)$   | ${r \over 2}{\rm \bf 32}+
                        ({4+n-r \over 2}){\rm \bf 32}^{\prime}+
                             (n+8){\rm \bf 12}$  | $2n+18$ \elt
 $D_6^{ns}$ | $SO(11)$  | $({n \over 2}+2){\rm \bf 32}+
                      (n+7){\rm \bf 11}$  | $3n+26$ \elt
 $A_5^{s}$ | $SU(6)$   | ${r \over 2} {\bf 20}+
     (16+r+2n) {\bf 6}+
           (2+n-r){\bf  15}$ |  $3n-r+21$ \elt
 $A_5^{ns}$  |  $Sp(3)$      |  $(16+2n+{3\over 2}r){\bf 6}  
+(n+1-r){\bf 14}
+\half r {\bf 14'}$   | $4n+23-2r$ \elt
 $A_4^{s}$ | $SU(5)$   | $(3n+16){\rm \bf 5}+(2+n){\rm \bf 10}$ |
        $5n+36$\elt
$A_2^{s}$ | $SU(3)_2$ | $(6n+r+34) {\rm \bf 3} + (r-2) {\rm \bf 6}  
+ (n+1-r)
{\rm \bf 8}$ | $4n+22-r$
\endtable
\vfill
\eject

Diagram 1: Higgs Tree
\bigskip
$$\matrix{SU(3)_2 & \leftarrow &Sp(3) & \leftarrow & \, & SU(6)
&\leftarrow & SO(12) & \,& \,\cr
\, &\, & \, & \, & \, & \,  &\, & \downarrow  &
\,& \,\cr
\, &\, &\, & \, & \, & \downarrow  &\, & SO(11) &
\leftarrow & E_7\cr
\, &\, &\, & \, & \, & \,  &\, & \downarrow  &
\,& \downarrow\cr
\downarrow &\, &\downarrow & \, & \, & SU(5)  &\leftarrow &
SO(10) & \leftarrow & E_6\cr
\, &\, &\, & \, & \, & \,  &\, & \downarrow  & \,&
\downarrow\cr
\, &\, &\, & \, & \, & \,  &\, & SO(9)  &
\leftarrow & F_4\cr
\downarrow &\, &\downarrow & \, & \, & \downarrow  &\, &
\downarrow \cr
\, &\, &\, & \, & \, & \,  &\, & SO(8) \cr
\, &\, &\, & \, & \, & \,  &\, & \downarrow \cr
\downarrow &\, &Sp(2) & \leftarrow & \, & SU(4)  &
\leftarrow & SO(7) \cr
\, &\, &\, & \, & \, & \downarrow  &\, & \downarrow \cr
\, &\, &\downarrow & \, & \, & SU(3)  &\leftarrow & G_2 \cr
\, &\, &\, & \, & \, & \downarrow & \,\cr
SU(2)_2 &\leftarrow &SO(4)       & \rightarrow &
\, &SU(2) & \,\cr}$$

\bigskip

\subsec{Unbroken $E_7$ gauge symmetry}

There is a subspace of the Higgs moduli space with unbroken $E_7$ when
the $12+n$ instantons are in commutant $H=SU(2)$.  The dimension of
this subspace, according to \ine, is $2n+21$.  In addition to these
neutral hypermultiplets, it follows from \nmatteri\ that there are
$(n+8)$ ${1\over 2}$-hypermultiplets in the ${\rm \bf 56}$ of $E_7$.
The codimension of this space of enhanced $E_7$ is given by the Higgs
mechanism to be $(n+8)(\half)(56)-133=28n+91$, leading to the expected
total dimension of $30n+112$.

Consider now F-theory with an $E_7$ geometric singularity.  According
to table 1, this is the case when ${\rm ord}(f) \geq 3$, and ${\rm
ord}(g) > 5$.  The F-theory moduli associated with the $E_8$ with
$12+n$ instantons satisfying these conditions are the terms ${\rm
ord}(f)=i=3$ and ${\rm ord}(g)=j=5$ in \expan, i.e.  the terms
$f_{8+n}$ and $g_{12+n}$.  The number of moduli associated with these
terms, subtracting one, as always, to account for the rescaling of
$z_1$ mentioned earlier, is $(13+n)+(9+n)-1=2n+21$.  This is exactly
the dimension found above for enhanced $E_7$ gauge symmetry, which is
a strong check of the proposed F-theory/heterotic duality.

The $E_7$ matter can be seen by considering the discriminant on the
$E_7$ locus:
\eqn\esevd{\Delta =  z_1 ^9 \big(4f_{8+n}^3 (z_2)+o(z_1) \big).}
The zero locus of the discriminant consists of the $z_1=0$ locus and
the other branch, which intersects the $z_1=0$ locus at $(n+8)$
points. From the point of view of type IIB string theory, the $z_1=0$
locus describes a 7-brane with $E_7$ vector fields propagating inside
its world volume.  The other branch has an interpretation as $(n+8)$
7-branes intersecting the 7-brane located at $z_1=0$.  We see a nice
match between extra zeroes of the discriminant, which coincide with
the zeroes of $f_{8+n}$, and the number of $\half {\bf 56}$
hypermultiplets.  We are thus led to conclude that each charged
${1\over 2}$-hypermultiplet is localized at a zero of $f_{8+n}$ \mv.

\subsec{Unbroken $E_6$ gauge symmetry}

There is a subspace of the moduli space with unbroken $E_6$ gauge
symmetry when the $12+n$ instantons of the heterotic theory are
embedded in commutant $SU(3)\subset E_8$.  Starting from the theory of
the previous subsection, relaxing the instantons from commutant
$SU(2)$ to $SU(3)$ corresponds to Higgs breaking of $E_7$ to $E_6$ by
giving an expectation value to two fields in the ${\bf 56}$.  It
follows either from the massless matter of the previous subsection and
the Higgs mechanism or from \ine\ and
\nmatteri\ applied to the commutant $SU(3)$ that the dimension
of the $E_6$ locus is $3n+28$ and that there are $n+6$ matter
hypermultiplets in the ${\bf 27}$ of $E_6$.

Let us compare this with the dimension of the locus in the F-theory
moduli space with an $E_6$ singularity.  To obtain an $E_6$
singularity, using table 2 and putting the equation in the Weierstrass
form, one finds that one has to first relax the restriction on $g(z_1,
z_2)$ in \expan, allowing a term $g_{2n+12}(z_2)$ in \expan\ (${\rm
ord}(g)=4$).  So the moduli of an $E_6$ singularity correspond to the
terms in $f_{n+8}$, $g_{n+12}$ and $g_{2n+12}$.
However, to actually obtain $E_6$ gauge symmetry, one
finds that the $E_6$ singularity in the fiber should not be generic,
but should be of the `split' form in table 2.  This implies that
there are $(n+6)$ constraints on the coefficients of $g_{2n+12}(z_2)$,
namely it should have double zeroes: $g_{2n+12}(z_2) =q_{n+6}^2
(z_2)$.  Counting the number of moduli in $q_{n+6}$, $g_{n+12}$, and
$f_{n+8}$, we find $n+7+n+13+n+9-1=3n+28$, in exact agreement with the
expected number from the heterotic side.

Further, we can again see the charged matter content from the zeroes of
the discriminant:
\eqn\esixd{\Delta=z_1 ^8 (27 q_{n+6} ^4 + o(z_1)).}
The $n+6$ {\bf 27}s of $E_6$ are localized at the zeroes of
$q_{n+6}$.

\subsec{Unbroken $F_4$}

The heterotic theory has an unbroken $F_4$ gauge symmetry when the
$12+n$ instantons lie in commutant $H=G_2\subset E_8$.  This theory
can be obtained from that of the previous subsection by giving an
expectation value to one of the fields in the ${\bf 27}$.  It follows
either from the Higgs mechanism or from \ine\ and \nmatteri\ applied
to the commutant $F_4$ that there are $4n+34$ neutral hypermultiplet
moduli and $n+5$ hypermultiplets in the ${\bf 26}$ of $F_4$.

Using Tate's algorithm, we find that $F_4$ gauge symmetry corresponds
to the generic $E_6$ singularity, relaxing the condition on
$g_{2n+12}$ found in the previous subsection.  This
relaxing, corresponding to splitting the double zeroes of $g_{2n+12}$
in the $E_6$ case, must have the interpretation of breaking $E_6$ to
$F_4$ by giving expectation values to the $F_4$ singlet components of
the ${\rm \bf 27}$s of $E_6$ (${\rm \bf 27}\rightarrow{\rm \bf
26}+{\rm \bf 1}$). Counting the dimension of the generic $E_6$
singularity, corresponding to the terms in $f_{n+8}$, $g_{n+12}$, and
$g_{2n+12}$, we find $n+9+n+13+2n+13-1=4n+34$, in precise agreement
with the expected dimension for unbroken $F_4$ gauge symmetry.

In this example, it is difficult to localize the expected $n+5$
hypermultiplets in the $26$ of $F_4$ at the extra zeroes of the
discriminant.  This seems to be the case for all the non-simply laced
gauge symmetries we will find below as well.

\subsec{Unbroken $SO(11)$}

There is a locus of unbroken $SO(11)$, which can be obtained by
starting from the above $E_7$ locus and giving expectation values
to components of two $\bf {56}$s.  
The massless hypermultiplet content consists of $3n+26$
singlet moduli, $n+7$ hypermultiplets in the ${\bf 11}$, and $\half
(n+4)$ in the ${\bf 32}$; this follows from the $E_7$ spectrum and
the Higgs mechanism or from \ine\ and \nmatteri\ applied to the
commutant, which is $SO(5)$.

In the F-theory this should correspond to compactification on an
elliptic Calabi-Yau with a {\it generic} $D_6$ singularity.  The
condition for such a singularity is that ${\rm ord}(f)=2$, ${\rm
ord}(g) \geq 3$ (or the signs $=,~\geq$ being permuted), with
coefficients chosen to cancel the $z_{1}^{6}$ and $z_1^7$ terms in
$\Delta$.  This requires $f_{2n+8}\sim s_{n+4}^2$, $g_{3n+12}\sim
s_{n+4}^3$, and $g_{2n+12}\sim f_{n+8}s_{n+12}$.  So such a manifold
is specified by independent functions
$$g_{12+n},~~f_{8+n},~~s_{4+n},$$ as follows immediately from Tate's
algorithm.  The resulting locus has dimension $3n+26$, which exactly
agrees with the heterotic string prediction.

\subsec{Unbroken $SO(10)$}

There is a locus of unbroken $SO(10)$ which can be obtained from the
$E_6$ of subsec. 4.2 by giving expectation values to two of the ${\bf
27}$ hypermultiplet flavors or from the $SO(11)$ of subsec. 4.4 by
giving an expectation value to one of the ${\bf 11}$ flavors.  It
follows from the Higgs mechanism applied to either route or from \ine\
and \nmatteri\ applied to the $SO(10)$ commutant, which is $SU(4)$,
that there are $4n+33$ neutral hypermultiplets, $(n+4)$ ${\bf16}$s,
and $n+6$ ${\bf10}$s.

On the F-theory side, according to table 2, to obtain $SO(10)$ gauge
symmetry the fiber should have a `split' $D_5$ singularity.  The
condition for a non-split $D_5$ singularity is that ${\rm ord}(f)=2$,
${\rm ord}(g) \geq 3$ (or the signs $=,~\geq$ being permuted), with
coefficients chosen to cancel the $z_{1}^{6}$ term in $\Delta$.  That
means that the polynomials $g_{3n+12}$ and $f_{2n+8}$ should be
related via $g_{3n+12} \sim h_{n+4}^3$ and $f_{2n+8} \sim
h_{n+4}^{2}$.  The condition of `splitness' in table 2 implies that
in addition we should take $g_{12+2n}=q_{n+6}^2- f_{8+n} h_{4+n}$.  So
the moduli of the split $D_5$ singularity correspond to the terms in
$h_{n+4}$, $q_{n+6}$, $g_{n+12}$, and $f_{n+8}$, for a total of
$4n+33$ hypermultiplet moduli, exactly as expected from the heterotic
side.  The discriminant on this locus is
\eqn\soten{\Delta = z_{1}^{7} h_{4+n} ^{3}
(q_{6+n}  ^2+o(z_1)).}
In this example, we can localize the matter at the extra zeroes of the
discriminant.  The $n+4$ ${\bf16}$s are localized at the zeroes of $h$
and the $n+6$ ${\bf10}$s are localized at the zeroes of $q_{n+6}$.

\subsec{Unbroken $SO(9)$}

The locus of unbroken $SO(9)$ gauge symmetry is reached by starting
either from the $F_4$ of subsec. 4.3 and giving an expectation value
to a field in the ${\bf 26}$ or starting from the $SO(10)$ of the
previous subsection and giving an expectation value to a field in the
${\bf 10}$.  The massless matter content is obtained either by the
Higgs mechanism applied to either route or by applying \ine\ and
\nmatteri\ to the commutant $SO(7)$:  There are $5n+39$ neutral
hypermultiplets, giving the dimension of the $SO(9)$ locus, $n+5$
hypermultiplets in the ${\bf 9}$ and $n+4$ in the ${\bf 16}$.

On the F-theory side, we see from table 2 that a manifold with a
generic (`nonsplit') $D_5$ singularity should yield $SO(9)$ gauge
group.  Relaxing the ``split'' condition of the previous subsection,
means that the moduli are now the terms in $h_{n+4}$, $g_{2n+12}$,
$g_{n+12}$, and $f_{n+8}$, for a total of $5n+39$, exactly as expected
{}from the heterotic side.

\subsec{Unbroken $SO(8)$}

The locus of unbroken $SO(8)$ is obtained from the $SO(9)$ above by
giving an expectation value to a field in the ${\bf 9}$.  The massless
matter content is obtained either from the Higgs mechanism or from
\ine\ and \nmatteri\ applied to the commutant, which is also $SO(8)$.
There are $6n+44$ hypermultiplets which are $SO(8)$ singlets, giving
the dimension of the $SO(8)$ locus, and $n+4$ hypermultiplets in the
$( {\rm \bf 8}_v+ {\rm \bf 8}_s+ {\rm \bf 8}_c)$.

Tate's algorithm implies that the F-theory side should be compactified
on an elliptic Calabi-Yau with a $D_4$ singular fiber satisfying
certain additional restrictions, as discussed after table 2.  These
restrictions tell us that $f_{2n + 8} (z_2)$ and $g_{3n +12} (z_2)$
only contribute $2n+10$ independent parameters instead of $5n+22$ --
they can be parameterized as $f_{2n + 8} (z_2)\sim h_{n+4}^2$ and
$g_{3n +12} (z_2)\sim q_{n+4}^3$.  This is a relaxation of the above
$D_5$ singularity, for which $h_{n+4}=q_{n+4}$.  Counting moduli, we
have the terms in $q_{n+4}$ in addition to those of the previous
subsection, leading to a total of $6n+44$ exactly as expected on the
heterotic side.  The discriminant in this parameterization can be
written as follows $$\Delta=z_1 ^6 \big( (h^2 _{n+4}+q^2 _{n+4})(h^2
_{n+4}+\omega q^2_{n+4}) (h^2 _{n+4} +
\omega^2 q^2 _{n+4}) + o(z_1)  \big)~,$$
where $\omega^3=1$.   The zeroes of each bracket correspond
to the charged matter multiplets.   There is a $Z_3$ symmetry $q
\rightarrow
\omega q$
exchanging the various factors in the discriminant -- this presumably
is a consequence of $SO(8)$ triality.  It is quite natural to
associate the $n+4$ matter fields in each of the three eight
dimensional representations of $SO(8)$ with each of the factors in the
discriminant.

\subsec{Unbroken $SO(7)$}

Continuing from the $SO(8)$ above, the Higgs mechanism implies that
the $SO(7)$ locus has $7n+48$ singlet hypermultiplets, $n+3$ in the
${\bf 7}$, and $2n+8$ in the ${\bf 8}$.  This massless spectrum can
also be obtained from \ine\ and \nmatteri\ applied to the commutant
$SO(9)$.

On the F-theory side, unbroken $SO(7)$ again corresponds to a $D_4$
singularity but with a different restriction, namely $g_{3n
+12}=f_{2n+8}q_{n+4}$, which follows from table 2.  The moduli of this
locus thus correspond to the terms in $f_{2n+8}$, $q_{n+4}$,
$f_{n+8}$, $g_{2n+12}$, and $g_{n+12}$, for a total of $7n +48$,
exactly as expected from the heterotic side.  The discriminant locus
is given by the zeroes of $$\Delta=z_1 ^6 \big(f_{2n+8} q^2_{n+4}
+o(z_1)
\big)~.$$ The zeroes of $f_{2n+8}$ correspond to the spinors, while
there is no such simple statement for vectors--as in all the examples
of non-simply laced groups, the origin of the matter does not seem to
be completely localized.

\subsec{Unbroken $G_2$}

There is a locus of unbroken $G_2$, which is obtained from the $SO(7)$
locus above by giving an expectation value to a hypermultiplet in the
${\bf 8}$.  The massless hypermultiplet content consists of $9n+56$
singlet moduli and $(3n+10)$ hypermultiplets in the ${\bf 7}$ of
$G_2$.  This spectrum is obtained either from the above $SO(7)$
spectrum and the Higgs mechanism or from \ine\ and \nmatteri\ applied
to the commutant $F_4$.

In F-theory, $G_2$ finally corresponds to {\it generic} $D_4$
singularity.  Compared to $SO(7)$, this corresponds to trading the
terms in $q_{n+4}$ for terms in $g_{3n+12}$, introducing $2n+8$ extra
moduli, for a total of $9n+56$ moduli, exactly as expected from the
heterotic side.  There are $(3n+10)$ ${\bf 7}$s which again are not
localizable in any obvious way.

\subsec{Unbroken $SU(4)$}

The locus of unbroken $SU(4)\cong SO(6)$ is obtained by starting from
the unbroken $SO(7)$ locus discussed above and giving an expectation
value to a field in the ${\bf 7}$.  The massless hypermultiplet
content consists of $8n+51$ singlet moduli, $n+2$ hypermultiplets in
the ${\bf 6}$, and $4n+16$ in the ${\bf 4}$; this follows from the
$SO(7)$ spectrum and the Higgs mechanism or from \ine\ and \nmatteri\
applied to the commutant, which is $SO(10)$.

It follows from table 2 that in order to get $SU(4)$ gauge symmetry we
need to take a compactification manifold with a `split' $A_3$ fiber
singularity.  A generic $A_3$ singularity implies various relations
between the polynomials in \expan: $$f_{8+4n}~\sim
{}~h_{4+2n}^{2},~~~f_{8+3n}~\sim ~h_{4+2n}H_{4+n}$$ $$g_{12+6n}
{}~=~h_{4+2n}^{3},~~~g_{12+5n} ~=~ -h_{4+2n}^{2}H_{4+n}$$ $$g_{12+4n}
{}~=~-f_{8+2n}h_{4+2n} + {1\over 12} h_{4+2n}H_{4+n}^{2}$$ $$g_{12+3n}
{}~=~{1\over 216}H_{4+n}^{3} + {1\over 6}f_{8+2n}H_{4+n} -
f_{8+n}h_{4+2n}.$$ To get the $SU(4)$ enhanced gauge
symmetry one needs a `split' $A_3$ singularity which implies that one
must impose the additional constraint $$h_{2n+4}=h_{n+2}^2.$$  The
moduli of the $SU(4)$ locus thus correspond to the terms in $h_{n+2}$,
$H_{n+4}$, $f_{2n+8}$, $f_{n+8}$, $g_{2n+12}$, and $g_{n+12}$ for a
total of $8n+51$, exactly as expected on the heterotic side.
The discriminant on this locus is equal to
\eqn\sufour{\Delta=z_1 ^4
\big(  h_{2+n} ^2  P_{16+4n} + o(z_1)  \big)~,}
where $P_{16+4n}$ is some polynomial constructed from the $f$s and
$g$s. The $h_{2+n}$ factor in $\Delta$ is responsible for the presence
of $n+2$ antisymmetric tensors in the $SU(4)$ theory, while the zeroes
of $P_{16+4n}$ yield $4n+16$ $\bf{4}$s.  Just as in the case of all
other simply laced groups, the matter seems localized.

\subsec{Unbroken $SO(5)$}

The locus of unbroken $SO(5)\cong Sp(2)$ can be obtained by starting
{}from the unbroken $SU(4)$ locus discussed above and giving an
expectation value to a field in the ${\bf 6}$.  The massless
hypermultiplet content consists of $9n+53$ singlet moduli, $n+1$
hypermultiplets in the ${\bf 5}$, and $4n+16$ in the ${\bf 4}$; this
follows from the $SU(4)$ spectrum and the Higgs mechanism or from
\ine\ and \nmatteri\ applied to the commutant, which is $SO(11)$.

It follows from table 2 that on the F-theory side one obtains $SO(5)$
gauge symmetry from the presence of a $\it generic$ $A_3$ singularity.
Starting from the split $A_3$ singularity this corresponds to relaxing
the condition $h_{2n+4}=h_{n+2}^2$.  The moduli thus correspond to the
terms of the previous subsection but with $h_{n+2}$ traded for
$h_{2n+4}$, yielding $n+2$ additional moduli for a total of $9n+53$ --
exactly as expected from the heterotic side.

\subsec{Unbroken $SO(4)$}

There is a locus of unbroken $SO(4)\cong SU(2)\times SU(2)$, which can
be reached from the $SO(5)$ locus discussed above by giving an
expectation value to a field in the ${\bf 5}$.  The massless
hypermultiplet content consists of $10n+54$ singlet moduli, $n$
hypermultiplets in the $({\bf 2}, {\bf 2})$, and $4n+16$ in the $({\bf
1}, {\bf 2})+({\bf 2}, {\bf 1})$; this follows from the $SO(5)$
spectrum and the Higgs mechanism or from \ine\ and \nmatteri\ applied
to the commutant, which is $SO(12)$.

In F-theory, $SO(4)$ appears when the discriminant locus has {\it two}
irreducible components of $D_u=D_v+nD_s$ (following the notation of
\mv) with an $A_1$ singularity on each. The codimension of this
configuration is
\foot{This codimension can be obtained by
applying ``the deficit argument''
introduced in section 6.} $2(12n+29)-4n=20n+58$, which agrees with
that expected from the heterotic side (recall the total dimension for
the bundle associated with this $E_8$ is $30n+112$). If $n=0$, this
corresponds to two independent theories. However for $n>0$ these two
components will intersect at $D_u^2=n$ points leading to $n$
hypermultiplets in the $({\bf 2},{\bf 2})$ \bsvo, agreeing with the
expected result of the heterotic side.

\subsec{Unbroken $SU(3)$}

The locus of unbroken $SU(3)$ can be obtained by starting either from
the unbroken $SU(4)$ locus and giving an expectation value to two
hypermultiplets in the ${\bf 4}$ or by starting from the $G_2$ locus
and giving an expectation value to a ${\bf 7}$ hypermultiplet.  The
massless hypermultiplet content consists of $12n+66$ singlet moduli
and $(6n+18)$ hypermultiplets in the ${\bf 3}$.  This
follows from either the $SU(4)$ or $G_2$ matter content discussed
above and the Higgs mechanism or from \ine\ and \nmatteri\ applied to
the commutant, which is $E_6$.

It follows from table 2 that a $\it split$ $A_2$ singularity must
yield $SU(3)$ gauge symmetry in F-theory.  The conditions on the
polynomials for such a split $A_2$ singularity are

\eqn\suthree{\eqalign{g_{12+6n} & \sim  h_{2+n}^{6}~~,~~~
                         f_{8+4n} \sim h_{2+n}^4~~,~~~
                         g_{12+5n}=-Q_{6+2n}h_{2+n}^3~~,~~~  \cr
                          f_{8+3n}& =h_{2+n}Q_{6+2n}~~,~~~
                          g_{12+4n}=-h_{2+n}^{2}f_{8+2n}  +
{1\over 12}Q_{6+2n}^{2}. }}
The moduli of the $SU(3)$ locus thus correspond to the terms in
$h_{n+2}$, $Q_{2n+6}$, $f_{2n+8}$, $f_{n+8}$, $g_{3n+12}$,
$g_{2n+12}$, and $g_{n+12}$, for a total of $12n+66$ exactly as
expected above.
The discriminant on this locus is given by
\eqn\suthree{\Delta = z_{1}^{3} h_{n+2}^{4} P_{16+5n}.}
Just as in the $SU(4)$ case, it is natural to associate $(n+2)$
antisymmetric tensors (which are the same as fundamentals for $SU(3)$)
to the zeroes of $h_{n+2}$ and $5n+16$ more hypermultiplets of
$\bf{3}$s to the zeroes of $P_{16+5n}$, for a total of $6n+18$
$\bf{3}$s, in agreement with the heterotic side.

\subsec{Unbroken $SU(2)$}

The locus of unbroken $SU(2)$ can be obtained by starting either from
the unbroken $SU(3)$ locus discussed above and giving an expectation
value to two hypermultiplets in the ${\bf 3}$, or by starting from the
unbroken $SO(4)$ locus discussed above and giving an expectation value
to two fields in the $({\bf 1}, {\bf 2})$ (or $({\bf 2}, {\bf 1})$).
The massless hypermultiplet content consists of $18n+83$ singlet
moduli, and $6n+16$ hypermultiplets in the ${\bf 2}$, this follows
{}from the above $SU(3)$ or $SO(4)$ matter content and the Higgs
mechanism or {}from \ine\ and \nmatteri\ applied to the commutant,
which is $E_7$.

A $\it generic$ $A_1$ singularity yields an $SU(2)$ gauge group.
The condition that the discriminant  $\Delta$ and its derivative
vanish at $z_1=0$ implies that the polynomials $f$
and  $g$ must satisfy the relations
$$g_{12+6n}  \sim h_{4+2n}^3 ~~,~~~
                         f_{8+4n} \sim h_{4+2n}^2~~,~~~
        g_{12+5n}=-f_{8+3n}h_{4+2n}.$$
The moduli thus correspond to the terms in  $h_{2n+4}$, $f_{3n+8}$,
$f_{2n+8}$, $f_{n+8}$, $g_{4n+12}$, $g_{3n+12}$, $g_{2n+12}$, and
$g_{n+12}$ for a total of $18n+83$, exactly as expected from the
heterotic side.
The discriminant on the $SU(2)$ locus is equal to
\eqn\sutwo{\Delta = z_1 ^2 \big( h_{4+2n} ^2
P_{6n+16}  +o(z_1) \big)~.}
The zeroes of $h_{2n+4}$ give $(n+2)$ antisymmetric tensors (which are
singlets for $SU(2)$), while the $(6n+16)$ doublets of $SU(2)$ are
localized at the zeroes of $P_{6n+16}$.

\subsec{Unbroken $SU(2)_2$}

There is an unbroken, level two $SU(2)_2$ locus (for $n>0$) which can
be obtained {}from the $SO(4)$ locus discussed above by giving an
expectation to a field in the $(2,2)$.  It follows from the $SO(4)$
spectrum and the Higgs mechanism that the dimension of the $SU(2)_2$
locus is $11n+54$ hypermultiplet moduli and there are
$(8n+32)$ hypermultiplets in the ${\bf 2}$ and $(n-1)$ in the ${\bf
3}$.  This spectrum also follows from \ine\ and \nmatteri\ applied to
the commutant, which is $SO(13)$.

In F-theory one obtains $SU(2)_2$ by smoothing a configuration which
yields $SO(4)$. We remind the reader that $SO(4)$ appears when the
discriminant has two rational components $D_u$ (in the notation of
\mv) which necessarily intersect each other in $n$ points. Such a
configuration has codimension $20n+58$. Smoothing introduces $n$
complex parameters leading to codimension $19n+58$ for $SU(2)_2$ in
agreement with the Higgs mechanism. The smooth curve ${\cal C}$ has
genus $(n-1)$ which is the number of adjoints ${\bf 3}$ in agreement
with the topological theory arguments \kmp\ briefly discussed below in
section 6.

\subsec{Unbroken $SO(12)$}

The heterotic theory has an unbroken $SO(12)$ when the $12+n$
instantons are embedded in the commutant $SO(4)\cong SU(2)\times
SU(2)$.  A new feature here is that there are different $SO(12)$ loci
corresponding to the different choices of how the $12+n$ instantons
are distributed in the two $SU(2)$ factors.  Because it follows from
\nmatteri\ that each $SU(2)$ must have at least four instantons, we
will parameterize the choices by putting $4+r$ instantons in the first
$SU(2)$ and $8+n-r$ in the second, with $r=0\dots n+4$.  It follows
{}from the obvious generalization of \ine\ and \nmatteri\ for
commutant $H=SU(2)\times SU(2)$ with these instanton numbers that
there are $2n+18$ singlet hypermultiplet moduli, giving the dimension
of each $SO(12)$ locus, $n+8$ hypermultiplets in the ${\bf 12}$,
$\half r$ in the ${\bf 32}$, and $\half(4+n-r)$ in the ${\bf 32'}$.

On the F-theory side, table 2 (and the discussion after it)
says that we will find $SO(12)$ gauge
symmetry enhancement by starting with a $D_6$ fiber singularity and
imposing some additional restrictions.  The generic $D_6$ singularity
requires
\eqn\dsix{f_{8+2n} =
s_{n+4}^{2},\ g_{3n+12} = s_{n+4}^{3},\ g_{12+2n} =- s_{n+4}f_{n+8}.}
Working through Tate's algorithm as described in \S3, we see that
there are different ways we can further restrict to obtain $SO(12)$
gauge symmetry, parameterized by a single integer $r$.  This is in
agreement with the freedom of choosing how to divide the $12+n$
instantons between the two $SU(2)$s in $E_8$ on the heterotic side.
The basic condition (in the notation of \S3) is that the polynomial $$
a_{2,1}X^{2} ~+~a_{4,3}X ~+~a_{6,5}$$ should factorize as
$$ (p_{4+n-r}X + q_{8+n-r})~(t_{r}X + u_{r+4})~.$$ Transforming to
Weierstrass form, this implies relations of the form
\eqn\further{s_{n+4} ~\sim~p_{4+n-r}t_{r},
{}~~g_{n+12} ~\sim~ q_{8+n-r}u_{r+4},~~ f_{8+n} ~\sim~q_{8+n-r}t_{r}
{}~+~p_{4+n-r}u_{r+4}~.}  The parameters in
$p_{4+n-r},q_{8+n-r},t_r,u_{r+4}$, after subtracting
two degrees of freedom which can be absorbed in reparameterizations,
give us a $2n+18$ dimensional $SO(12)$ locus, as expected from the
heterotic side.

The discriminant looks like
$$\Delta ~\sim~z_{1}^{8}p_{4+n-r}^{2}t_{r}^{2}P_{8+n}^{2}.$$
We can associate the $r$ 1/2 ${\bf 32}$s with the zeroes of
$t$, the $(4+n-r)$ 1/2 ${\bf 32^\prime}$s with the zeroes of
$p$, and the vectors with the zeroes of $P_{8+n}$.

Note that giving an expectation value to a ${\bf 12}$ hypermultiplet
Higgses to the $SO(11)$ locus discussed above.  The distinction
between the different $SO(12)$ loci, labeled above by $r$, is lost
upon Higgsing to $SO(11)$, corresponding to the fact that the
commutant is enhanced to $SO(5)$, which is simple.

\subsec{Unbroken $SU(6)$}

There are different loci of unbroken $SU(6)$, corresponding to the
different ways of distributing the $12+n$ instantons among the two
factors in the commutant, which is $SU(2)\times SU(3)$.  The choices
can again be labeled by putting $4+r$ instantons in the first factor
and $8+n-r$ in the second, $0\leq r\leq n+2$.  These $SU(6)$ loci can
be obtained from the above $SO(12)$ loci by the Higgs mechanism,
giving an expectation value to the appropriate component of a ${\bf
32'}$.  The massless hypermultiplet content consists of $3n+21-r$
singlet moduli, $(16+2n+r)$ hypermultiplets in the ${\bf 6}$,
$2+n-r$ in the ${\bf 15}$ and
$\half r$ in the {\bf 20}.

In F-theory, one obtains $SU(6)$ gauge group from a split $A_5$
singularity.  It follows from Tate's algorithm that the most generic
such singularity is specified by polynomials $$f_{4+2n} =
h_{2+n}^{2},~~f_{8+n},~~s_{4+n},~~f_{4}.$$ This is the same as generic
$A_5$ except for the constraint that
\eqn\susix{f_{2n+4}(z_2)=h_{n+2}^2(z_2).}
The discriminant locus looks like
\eqn\sixdisc{\Delta = z_{1}^{6} h_{n+2}^{4} P_{2n+16}}
This locus has dimension $3n+21$ and corresponds to the
$r=0$ case of the heterotic string, with the $2n+16$ ${\bf 6}$s
localized at the zeroes of $P_{2n+16}$ and
the $2+n$ ${\bf 15}$s localized
at the zeroes of $q_{n+2}$.

Natural candidates for the F-theory duals to the heterotic theories
with $r\neq 0$ are the special `split' $A_5$ singularities
which satisfy the constraints \further\ of the $r$th $SO(12)$ locus
and in which
\eqn\spsplit{h_{n+2} ~=~t_{r}\tilde h_{2+n-r},}
with $t_r$ as in subsection 4.16.
The independent polynomials are $\tilde h_{2+n-r}, t_{r}$, $q_{8+n-r},
p_{4+n-r}$ and $u_{4+r}$.
Remembering to subtract the $\it two$ reparameterizations, we
see that such a theory occurs at
dimension $2n+18 + n+3-r$ = $3n+21-r$, in agreement with the
series of heterotic $SU(6)$ theories parameterized by $r$.

\subsec{Unbroken $SU(5)$}

Unbroken $SU(5)$ can be obtained either from the $SO(10)$ by Higgsing
with a ${\bf 16}$ or from $SU(6)$ by Higgsing with two fundamentals.
(All of the $SU(6)$ theories, labeled by $r$, Higgs to the same $SU(5)$
theory.)  The massless matter spectrum consists of $5n+36$
hypermultiplet moduli, $3n+16$ in the ${\bf 5}$ and $n+2$ in the ${\bf
10}$.  This spectrum can be obtained from that of $SO(10)$ or $SU(6)$
and the Higgs mechanism, or by applying \ine\ and \nmatteri\ to the
commutant, which is also $SU(5)$.

It follows from table 2 that in F-theory one obtains $SU(5)$ gauge
group by compactifying on a manifold with a split $A_4$ singularity.
It follows from Tate's algorithm that such a singularity is given by
specifying five polynomials $h_{2+n}$, $H_{4+n}$, $q_{6+n}$,
$f_{8+n}$, $g_{12+n}$.  The other $f$s and $g$s are specified in terms
of these, e.g.  $$g_{12+6n}~~\sim ~~h_{2+n}^{6}, ~~f_{8+4n}~~ \sim~~
h_{2+n}^{4}, \dots$$ The dimension of the $SU(5)$ locus is thus
$5n+36$, exactly as expected from the heterotic side.  The
discriminant locus looks like $$\Delta \sim z_{1}^{5}
{}~h_{2+n}^{4}~P_{16+3n}~.$$ The $3n+16$ zeroes of $P$ yield the ${\bf
5}$s while the zeroes of $h_{2+n}$ correspond to the antisymmetric
tensor ${\bf 10}$s.

\subsec{Unbroken $Sp(3)$}

As above, there are theories labeled by an integer $r$ corresponding
to the different ways of distributing the $12+n$ instantons among the
factors in the commutant, which is $SU(2)\times G_2$.  These $Sp(3)$
loci can be obtained from $SU(6)$ by giving an expectation value to
one of the ${\bf 15}$ hypermultiplets.  The hypermultiplet content
consists of $4n+23-2r$ singlet moduli, $16+2n+{3\over 2}r$
hypermultiplets in the ${\bf 6}$, $n+1-r$ in the ${\bf 14}$, and
$\half r$ in the three index antisymmetric ${\bf 14'}$.

In F-theory, one can obtain this family of theories by taking the
$r$th $SU(6)$ theory of subsection 4.17 and relaxing the condition
\susix\ on
$f_{2n+4}$ to be $$f_{2n+4} = \tilde h_{2n+4-2r} t_{r}^{2}.$$ For
$r=0$, this is the generic $A_5$ singularity.  The independent
polynomials going into specifying the $Sp(3)$ theories are $\tilde
h_{2n+4-2r}, t_{r}, q_{8+n-r}, p_{4+n-r}$ and $u_{4+r}$.  The
dimension increases by $n+2-r$ with respect to the $r$th $SU(6)$
locus, hence it is given by $4n+ 23 - 2r$, in agreement with the
expected result from the heterotic side.

\subsec{Unbroken $SU(3)_2$}

The above $Sp(3)$ theory can be broken to a level $2$ $SU(3)$ by
giving an expectation value to the $14'$.  The dimension of the
$SU(3)_2$ locus is $4n+22-r$ and there are $34+6n+r$ hypermultiplets
in the ${\bf 3}$, $r-2$ in the ${\bf 6}$, and $n+1-r$ in the ${\bf
8}$.  This spectrum is obtained either from this Higgsing or by
applying \ine\ and \nmatteri\ to the commutant, which is $SU(3)\times
G_2$.

One can also obtain $SU(3)_2$ from the $SU(3)\times SU(3)$ theory in
the same way as one obtains $SU(2)_2$ from $SO(4)=SU(2)\times
SU(2)$. The theory with $SU(3)\times SU(3)$ appears at codimension
$2(18n+46)-9n=27n+92$ (this follows from the ``deficit argument''
discussed in sect. 6) when the discriminant has two $A_2$ components
$D_u$ intersecting in $n$ points. Such a theory has $n$
hypermultiplets in the $({\bf 3},{\bf 3})$ and $3n+18$ hypermultiplets
in the $({\bf 3},{\bf 1})+({\bf 1},{\bf 3})$. Smoothing $n+2-r$ double
points one adds $n+2-r$ parameters to end up with codimension
$26n+r+90$, exactly as expected from the heterotic side.  This
smoothing corresponds to Higgsing $n+2-r$ of the $n$ mixed $({\bf
3},{\bf 3})$ hypermultiplets. The remaining $r-2$ mixed
representations decompose as $({\bf 3})+({\bf 6})$ reproducing the
spectrum of $SU(3)_2$.  For $r=2$ the genus of the {\it smooth} curve
is $n-1$ in agreement with the number of adjoints ${\bf 8}$. For $r>2$
the resulting curve is not smooth, but it has exactly $n+1-r$
holomorphic 1-differentials which in the topological theory correspond
to adjoints. The remaining $r-2$ double points thus correspond to the
matter in the ${\bf 3} + {\bf 6}$.

\newsec{Small Instantons}

The $E_8 \times E_8$ heterotic string with instanton numbers $(16,8)$
embedded in the two $E_8$s has been conjectured \mv\ to be equivalent
to the $SO(32)$ heterotic string.  In fact, this has recently been
established via T-duality \ref\berketal{M. Berkooz et. al.,
``Anomalies, Dualities, and Topology of D=6, N=1 Superstring Vacua,''
hepth/9605184.}.  Note that both theories have generic unbroken
$SO(8)$ gauge symmetry; in the $E_8 \times E_8$ theory that is the
generic unbroken gauge symmetry associated with the $E_8$ with 8
instantons, while in the $SO(32)$ theory it is the generic unbroken
gauge group with 24 instantons.  The $SO(32)$ string is known to
develop an enhanced $Sp(1)\simeq SU(2)$ non-perturbative gauge
symmetry whenever an instanton collapses to zero size \witsi.  When
$k$ collapse at the same point in $K3$, an $Sp(k)$ gauge symmetry
develops non-perturbatively.  In light of the above correspondence, we
should be able to find these non-perturbative enhanced gauge
symmetries in the context of F-theory on the elliptic fibration over
$F_{4}$.  We will demonstrate in this section that this is so.

We have seen that the singularities of F-theory in $z_1$ correspond
to perturbative enhanced gauge symmetry of the heterotic theory.
Singularities of F-theory at $z_2 =const.$, on
the other hand,  should be interpreted as non-perturbative gauge
symmetries in the dual heterotic picture.  Therefore, we should
recover the results of
\witsi\ by studying degenerations of this sort in the $n=4$ model.

The case of a single small instanton leads to an enhanced
non-perturbative $Sp(1)$ and should thus correspond to an $A_{1}$
singularity located at a point on the base, say $z_{2} = 0$.  We
expand $f$ and $g$ as
\eqn\exp{f = w_{1}^{2} f_0 + w_{1}^{3} f_{4}(z_{2}) + \cdots,~~
g = w_{1}^{3} g_{0} + w_{1}^{4} g_{4}(z_{2}) + \cdots,}
where, for convenience, we expand in terms of $w_1=1/z_1$ around
$w_1=0$.  The discriminant takes the form
\eqn\disexpo{\Delta =w_1^6(D_0+w_1D_4(z_2)+\cdots).}
Note that our conditions for an unbroken $SO(8)$ perturbative gauge
symmetry are satisfied at $w_1=0$.  We now impose the condition for an
$A_{1}$ singularity, i.e. that $\Delta$ develops a double zero, ${\it
in}$ $z_{2}$.  This condition implies that $f$ and $g$ must satisfy
\eqn\aonecond{\eqalign{f(w_{1},0 )=3h^{2}(w_1),~&~g(w_1,0) =
-2h^{3}(w_1)\cr g^{\prime} (w_1,0) =& -hf^{\prime} (w_1,0)}} for some
function $h_3(w_1)$ of degree 3.  In this process, we trade the $7+ 10
+ 6 +9$ parameters in $f,g,f^\prime, g^\prime$ for the $4+6$
parameters in $h$ and $f^\prime$.  After subtracting one to account
for the arbitrary choice of the location of the singularity at
$z_2=0$, we see that the codimension for an enhanced {\it
non-perturbative} $Sp(1)$ is 21.

In addition to there being an enhanced {\it non-perturbative} gauge
symmetry, the {\it perturbative} gauge symmetry is also automatically
enhanced on the locus \aonecond.  To see that, note that the condition
that the discriminant \disexpo\ have a second order zero at $z_2=0$
requires the constant term $D_0$ to vanish, yielding a $D_5$
singularity at $w_1=0$.  Therefore, according to our dictionary,
the perturbative $SO(8)$ gauge symmetry is automatically extended to
$SO(9)$.  To summarize, we are finding the generic unbroken $SO(8)$
enhanced to an unbroken $SO(9)\times Sp(1)$, with $SO(9)$ perturbative
and $Sp(1)$ non-perturbative, at codimension $21$.

The above result is perfect for the conjectured equivalence of $n=4$
F-theory with the $SO(32)$ heterotic string on $K3$!  Shrinking one
instanton to zero size gives an enhanced non-perturbative $Sp(1)$
\witsi\ and the remaining 23 instantons can only break $SO(32)$
to $SO(9)$.  A single small instanton in this model is expected to
lead to $SO(9)\times SU(2)$ as the generic unbroken gauge group with
matter ${1\over 2}$ $({\bf 9}, {\bf 2})$ and ${23\over 2}$ $({\bf
1},{\bf 2})$ hypermultiplets.  The codimension for this enhanced gauge
symmetry is $21$, precisely as found above.

Consider, more generally, the condition for an enhanced
non-perturbative $Sp(k)$ in the $n=4$ F-theory.  We saw in sect. 4
that, for all $n$, a {\it perturbative} enhanced $Sp(k)$ gauge
symmetry corresponds to an $A_{2k-1}$ singularity in $z_1$
\foot{By induction, it can be shown that a
$A_{2k-1}$ singularity in $z_1$
occurs at codimension $-2nk^2+15nk+30k-n-1$.  This agrees with the
codimension for an enhanced
perturbative $Sp(k)$
with matter $16+2n(4-k)$ fundamentals and $n+1$ two-index antisymmetric
tensors.}.
Therefore, we should adjust the moduli so as to obtain an $A_{2k-1}$
singularity in $z_2$ to obtain an enhanced {\it non-perturbative}
$Sp(k)$ gauge symmetry.

To find the conditions and the codimension for an $A_{2k-1}$
singularity at $z_2=0$, note that the derivatives of the functions $f$
and $g$ in \exp\ with respect to $z_2$ have degree in $z_1$ given by
$deg( f^{(r)}(z_1,0))=6-[(r+3)/4]$ and
$deg(g^{(r)}(z_1,0))=9-[(r+3)/4]$, where $[\ ]$ denotes the integer
part.  It can thus be seen that an $A_{2k-1}$ singularity is enhanced
to an $A_{2k+1}$ singularity at codimension $22-k$. Iterating
this, the codimension for an $A_{2k-1}$
singularity at $z_2=0$, and thus a non-perturbative $Sp(k)$, is $\half
k(45-k)-1$.

As in the above $k=1$ case, the singularity at $z_1=\infty$ ($w_1=0$)
is also automatically enhanced, corresponding to a larger perturbative
gauge group.  It is easily seen in the first few cases that the
singularity at $w_1=0$ is precisely that which we found for a
perturbative $SO(8+k)$ gauge symmetry.  For example, an $A_{3}$
singularity in $z_2$ requires $D_{4}(z_2)=(const) z_{2}^{4}$ in
\disexpo\ which leads automatically to a $D_5$ singularity at
$w_1=0$ with the extra condition for $SO(10)$ gauge symmetry; an $A_5$
singularity in $z_2$
leads to a $D_6$ singularity at $w_1=0$, our condition for $SO(11)$
gauge symmetry; etc.  To summarize, we have an enhanced $SO(8+k)\times
Sp(k)$ gauge symmetry, with $SO(8+k)$ perturbative and $Sp(k)$
non-perturbative, at codimension $\half k(45-k)-1$.  The matter
content for this theory is hypermultiplets in the $\half ({\bf
8+k},{\bf 2k})$, ${1\over 2} (24-k) ({\bf 1},{\bf 2k})$, and $({\bf 1},{\bf
k(2k-1)-1})$.  The codimension for this enhanced gauge symmetry, by
the Higgs counting, is $\half k(45-k)-1$, agreeing with the
codimension found above for an $A_{2k-1}$ singularity in $z_2$.
Again, this enhanced gauge symmetry and matter content perfectly
agrees with the expected result, based on the considerations of
\witsi, for the generic unbroken gauge group and matter content for
$k$ small instantons of the $SO(32)$ heterotic theory at the same
point in $K3$!

\newsec{Mixing Perturbative and Non-Perturbative Gauge Symmetry}

\def \P1{{\bf P}^1}

In this section we focus on
the $n=0$ case, which corresponds to the symmetric $(12,12)$ heterotic
compactification on $K3$.
For $n=0$, the Hirzebruch surface $F_0=\P1\times\P1$ is a product of
two projective lines. The K\"ahler class decomposes as $k=k_1 a+k_2b$
where $k_1$ and $k_2$ are the areas of these $\P1$s and $a$ and $b$
are their dual 2-cocycles. The string coupling constant is given by
\mv
\eqn\scc{
{1\over \lambda^2}={\rm exp}(2\phi)={k_1\over k_2}.}  There is a
manifest ${\bf
Z}_2$ symmetry exchanging the factors, which was
interpreted in \refs{\mv,\aspgi} as the strong-weak coupling
duality of the $(12,12)$ heterotic theory proposed in \DMW.
This duality exchanges gauge symmetries which arise perturbatively in
the heterotic theory with ones which arise non-perturbatively.  In
this section we will discuss new enhanced gauge symmetries which are
mixtures of being perturbative and non-perturbative and are naturally
seen in $F$ theory.  Some cases of these mixed gauge symmetries were
also seen in the orientifold analysis of \GP.

In Section 4, we found the perturbative enhanced gauge symmetry of
the heterotic theory from
the singularities along $z_1=0$ (or more generally, the section
$z_{1}=P_{n}(z_{1})$).
In Section 5, we considered singularities along
$z_2=const$, finding the non-perturbative enhanced gauge symmetry
associated with small $SO(32)$ instantons.  In the present section, we
will consider singularities along a more general curve $\Sigma _{p,q}$
given by $F_{p,q}(z_1,z_2)=0$, where $(p,q)$ denotes the degree of the
polynomial in $z_1$ and $z_2$, respectively.
Once we have chosen to call the gauge symmetry corresponding to
$(1,0)$ ``perturbative,'' we should call the gauge symmetry
corresponding to $(0,1)$ ``non-perturbative''.  The more general
$(p,q)$ singularity corresponds to enhanced gauge symmetry which is a
mixture of perturbative and non-perturbative.

At this stage, it is useful to recall the constraint of six
dimensional anomaly factorization, discussed in \schwarz: the anomaly
polynomial should factorize as
\eqn\anomfact{I=(R^2-\sum _a u_aF_a^2)(R^2-\sum _a v_a F_a^2),}
where $a$ runs over the different gauge groups.  A given gauge group
and matter content thus has an associated $(u,v)$, which enter in the
gauge kinetic terms \refs{\sagnotti, \DMW}.  The level one
perturbative gauge groups and matter content associated with the
$(12,12)$ heterotic theory have $(u,v)=(2,0)$, while the
non-perturbative gauge groups have $(u,v)=(0,2)$ \DMW.  (Here we are
normalizing $(u,v)$ in
\anomfact\ by defining
$F_a^2$ to be the trace normalized by the index of the representation;
for example, $F_a^2=\tr F_a^2$ for $SU(N)$ and $\half \tr F_a^2$ for
$SO(N)$, with $\tr$ in the fundamental representation in both cases).

Consider first the case of a $\Sigma _{1,1}$ given by $z_1z_2=0$, with
$ A_1$ singularities along both $z_1=0$ and $z_2=0$.  With only the
$A_1$ singularity along $z_1=0$, we would expect a perturbative
enhanced $SU(2)_{(1,0)}$ with $16$ fundamental hypermultiplets and
$(u,v)=(2,0)$.  With only the $ A_1$ singularity along $z_2=0$, we
would expect a non-perturbative $SU(2)_{(0,1)}$, also with $16$
fundamental hypermultiplets, and $(u,v)=(0,2)$.  With both
singularities, we expect to find an enhanced $SU(2)_{(1,0)}\times
SU(2)_{(0,1)}$ with matter given by a single $({\bf 2},{\bf 2})$ field
and $14$ fields in the $({\bf 1},{\bf 2})+({\bf 2},{\bf 1})$.
Indeed, this matter content\foot{This theory and matter content can be
obtained from the $U(16)\times U(16)$ of \GP\ upon Higgsing (up to
$U(1)s$).  The perturbative $SU(2)$ comes {}from the $SO(32)$ 9 branes
upon Higgsing by Wilson lines and the non-perturbative $SU(2)$ is
associated with a small instanton \witsi\ or Type-1 5 brane.  The
$({\bf 2},{\bf 2})$ field comes from the $5-9$ sector.}  is the unique
solution of the above anomaly factorization condition for which
$SU(2)_{(1,0)}$ and $SU(2)_{(0,1)}$ are coupled, as they should be
because $z_1=0$ and $z_2=0$ intersect at a point, and which properly
reduces to $SU(2)_{(0,1)}$ or $SU(2)_{(1,0)}$ when the singularity
along $z_1$ or $z_2$ is smoothed.  The fact that there is a matter
field in the $({\bf 2}, {\bf 2})$ also follows from the intersecting
D-brane picture discussed in \bsvo.  The codimension for the enhanced
$SU(2)\times SU(2)$ with this matter content is $4+56-6=54$.

In F-theory, we interpret enhanced gauge symmetry as coming from
coinciding 7-branes wrapped around components of discriminant.  It is
natural that the gauge symmetry associated with $A_1$ singularities
along $z_1=0$ and $z_2=0$ should be two copies of the $SU(2)$ theory
which are coupled by a single $({\bf 2},{\bf 2})$ matter field,
corresponding to the intersection at a point.  Consider now the
codimension in F-theory for having $ A_1$ singularities along both
$z_1=0$ and $z_2=0$.  Near $z_1=z_2=0$, an $ A_1$ singularity of the
discriminant $\Delta (z_1,z_2)$ in either $z_1$ or $z_2$ is obtained
at codimension $29$, as in section 4.14.  Having an $ A_1$ singularity
in both $z_1$ and $z_2$ is almost just the sum of the two conditions,
but the vanishing of $\Delta$ to second order in both $z_1$ and $z_2$
kills the four terms with $\Delta \sim 1,z_1,z_2,z_1z_2$
twice\foot{This addition of the independent codimensions and
subtraction of the twice counted terms is the ``deficit argument''
referred to in the previous section.}.  So
the codimension is $29+29-4=54$, in agreement with the above expected
result!

The $SU(2)_{(1,0)}\times SU(2)_{(0,1)}$ theory can be Higgsed by
giving expectation values to the fields in the $({\bf 2},{\bf 1})$ or
$({\bf 1},{\bf 2})$, breaking the theory to the purely non-perturbative
$SU(2)_{(0,1)}$ or perturbative $SU(2)_{(1,0)}$ theories,
respectively, discussed before.  In terms of $F$ theory, such Higgsing
corresponds to smoothing the $ A_1$ singularity along $z_1$
or $z_2$.  Another possibility is to give an expectation value
to the $({\bf
2},{\bf 2})$ field, which breaks $SU(2)_{(1,0)}\times SU(2)_{(0,1)}$
to the diagonally embedded $SU(2)_{(1,1)}$ with $28$ fundamentals
(coming from the fourteen $({\bf 2},1)+(1,{\bf 2})$).  Because this
theory is obtained by Higgsing from one with factorized anomaly, it of
course also has a factorized anomaly.  The coefficients in
\anomfact\ are $(u,v)=(2,2)$.  In terms of $F$ theory, this
latter Higgsing corresponds to smoothing $\Sigma _{(1,1)}$ from
$z_1z_2=0$ to $z_1z_2=\epsilon$.  The codimension for an $
A_1$ singularity along this smoothed surface is $53$, which agrees
with the codimension for $SU(2)_{(1,1)}$ with $28$ doublets.

So F-theory predicts a new enhanced $SU(2)$ which is neither purely
perturbative nor purely non-perturbative.
It could not have been
seen from the arguments of \DMW\ because the ${\bf Z}_2$ strong-weak
duality
maps $(1,1)$ to itself.

We can extend the above analysis in two directions.  One is to
consider other singularities, which would produce other enhanced gauge
groups.  For example, consider $Sp(n)\times Sp(m)$, corresponding to
an $ A_{2n-1}$ singularity at $z_1=0$ and a $
A_{2m-1}$ singularity at $z_2=0$.  The matter content is a field in
the $({\bf 2n},{\bf 2m})$, $16-2m$ fields in the $({\bf 2n},{\bf 1})$,
$16-2m$ fields in the $({\bf 1},{\bf 2m})$, a field in the $({\bf
n(2n-1)-1},{\bf 1})$ and a field in the $({\bf 1,m(2m-1)-1})$.  The
codimension is $4nm+2n(16-2m)+2m(16-2n)+n(2n-1)-1
+m(2m-1)-1-n(2n+1)-m(2m+1)=30(n+m)- 4nm-2$.  In F-theory an
$ A_{2n-1}$ singularity has codimension $30n-1$ (as remarked
in footnote 6).  So an $ A_{2n-1}$ along $z_1=0$ and
$ A_{2m-1}$ along $z_2=0$ has codimension $30(n+m)-2-4nm$,
where the last term corresponds to the fact that the terms $a_{ij}$
with $\Delta \sim \sum _{i=0}^{2n-1}\sum
_{j=0}^{2m-1}a_{ij}z_1^iz_2^j$ were killed twice.  So the $F$ theory
gives the correct codimension.

Another extension of the above ideas is to more perturbative and (or)
non-perturbative gauge groups.  In F-theory it is clear that we can,
more generally, have a ``grid'' $\Sigma _{p,q}$, given by $\prod
_{i=1}^p(z_1-a_i)\prod _{j=1}^q(z_2-b_j)=0$, with $p$ singularities
along $z_1=a_i$, with $a_i$ constants and $q$ singularities along
$z_2=b_j$, with $b_j$ constants.  With $ A_1$ singularities
for each line of the grid, the enhanced gauge group is
$\prod_{i=1}^{p} SU(2)_{(1,0)}^{(i)}\times  
\prod_{j=1}^{q}SU(2)_{(0,1)}^{(j)}$
with $({\bf 2}^{(i)},{\bf 2}^{(j)})$ matter fields coupling each
$SU(2)_{(1,0)}^{(i)}$ to each $SU(2)_{(0,1)}^{(j)}$, corresponding to
the vertices of the grid, and $16-2q$ matter fields ${\bf 2}^{(i)}$
transforming transforming as fundamentals only under each of the
$SU(2)^{(i)}_{(1,0)}$ gauge groups and $16-2p$ matter fields ${\bf
2}^{(j)}$ transforming as fundamentals only under each of the
$SU(2)^{(j)}_{(0,1)}$ gauge groups.  It is easily verified that this
is the unique solution of the anomaly factorization equation for $p$
perturbative $SU(2)$s with $(u_i,v_i)=(2,0)$ and $q$ non-perturbative
$SU(2)$s with $(u_j,v_j)=(0,2)$.  The codimension for this enhanced
gauge group and matter content is
$4pq+2p(16-2q)+2q(16-2p)-3(p+q)=29(p+q)-4pq$.  This agrees with the
codimension computed in $F$ theory: Again, each of the $p+q$ $ A_1$
singularities occur at codimension $29$ but $4pq$ terms in $\Delta$
are killed twice.

By giving expectation values to the fields in the various $({\bf
2}^{(i)},{\bf 2}^{(j)})$, it is possible to Higgs to a variety of
different gauge groups which are neither purely perturbative nor
purely non-perturbative.  In F-theory this corresponds to smoothing
the various intersections of the ``grid'' as with $z_1z_2=0$ deformed
to $z_1z_2=\epsilon$.  This leads to more general diagrams of
intersecting singularities in the $z_1,z_2$ plane.  Generally, by this
Higgsing, we can get gauge groups $\prod _i SU(2)_{(p_i,q_i)}$ with
matter given by $p_iq_j+q_ip_j$ fields transforming as a $({\bf
2}^{(i)},{\bf 2}^{(j)})$ fundamental under both $SU(2)_{(p_i,q_i)}$
and $SU(2)_{(p_j,q_j)}$, $16(p_i+q_i)-4p_iq_i$ matter fields
transforming only as the fundamental ${\bf 2}^{(i)}$ under
$SU(2)_{(p_i,q_i)}$, and $(p_i-1)(q_i-1)$ matter fields transforming
only as an adjoint ${\bf 3}^{(i)}$ of $SU(2)_{(p_i,q_i)}$. \foot{The
basic mechanism for these results is illustrated by considering a
$SU(2)\times SU(2)$ with $N_v$ matter fields in the $({\bf 2},{\bf
2})$, $N_L$ matter fields in the $({\bf 2},{\bf 1})$ and $N_R$ matter
fields in the $({\bf 1},{\bf 2})$.  Giving an expectation value to one
of the $({\bf 2},{\bf 2})$ fields breaks to a diagonally embedded
$SU(2)$ with $N_v-1$ adjoints (plus some singlets) and $N_L+N_R$
fundamentals.  Iterating this leads to the various groups and matter
content described above.}  Because the $\prod _i SU(2)_{(p_i,q_i)}$
theory with this matter content was obtained by Higgsing from a theory
which satisfies the anomaly factorization condition, it of course also
satisfies this condition, as can be directly verified.  The gauge
group $SU(2)_{(p_i,q_i)}$ has $(u_i,v_i)=(2p_i,2q_i)$.  The above
matter content is the unique solution of the anomaly factorization
condition with only fundamentals and adjoints and with these values of
$(u,v)$.

As an extreme case, we can smooth all of the
intersections, Higgsing to a single $SU(2)$ with
$N_F=16(p+q)-4pq$ fundamentals and $N_A=(p-1)(q-1)$ adjoints.  This
theory has a factorized anomaly with
\eqn\uvpq{(u,v)=(2p,2q). }
The codimension for this $SU(2)$ is
$2[16(p+q)-4pq]+3(p-1)(q-1)-3=29(p+q)-5pq$.  The completely smoothed
grid corresponds to a surface $\Sigma$ of genus $(p-1)(q-1)$.  It is
interesting to note that, when $p=1$, this mixed
perturbative/non-perturbative spectrum for $E_8$ with $12$ instantons
coincides with the perturbative $SU(2)$ spectrum discussed in sect. 4
for $E_8$ with $12+2q$ instantons.

In $F$ theory, we can easily explain why $u/v=p/q$:  Following
\mv, the gauge coupling of the six-dimensional theory is proportional
to the integral
\eqn\gac{
{1\over g^2}=\int_C k=pk_1+qk_2\propto(pe^\phi + qe^{-\phi})} of the
K\"ahler class $k$ over the compact part $C$ of the 7-brane
world-volume. On the other hand, it follows from supersymmetry that
$g^{-2} \propto (ue^\phi +ve^{-\phi})$, so $u/v=p/q$.  To explain the
coefficient of $2$ in \uvpq, one would have to understand the anomaly
factorization property in terms of D-branes. Assuming \uvpq, we know
{\it a priori} that the above $SU(2)$ gauge theory appears when two
7-branes wrap around a smooth $(p,q)$ curve $C$.  On the other hand,
compactifying on $T^2$ down to 4 dimensions one can use the results of
\kmp\
 which predicts ${\rm
genus}(C)$ massless hypermultiplets in the adjoint representation in
this situation. With a little triumph, one notices that indeed ${\rm
genus}(C)=(p-1)(q-1)$, as it should be for consistency. Also, by the
above ``deficit argument'' the codimension of the corresponding locus
is $29(p+q)-5pq$, in agreement with the above Higgs mechanism
codimension.

Note that the above extensions are not unrelated. $SU(2)_{(m,n)}$ is
obtained by deforming away from a configuration of $m$ copies of
$SU(2)_{(1,0)}$ intersecting $n$ copies of $SU(2)_{(0,1)}$.  On the
other hand, bringing together all $n$ parallel 7-branes of such
configuration, one ends up with $Sp(n)\times Sp(m)$ theory on the
singular $(1,1)$ curve $zw=0$. When $n=m$ one can further break to
diagonal $Sp(n)$ by deforming to a smooth $zw=\epsilon$.

One can easily consider other types of singularities leading to a
variety of gauge groups.  For example, in the perturbative heterotic
theory in codimension 64 one finds an $SO(7)$ gauge theory with
$N_F=3$ hypermultiplets in $({\bf 7})$ and $N_S=8$ hypermultiplets in
$({\bf 8})$. This $SO(7)$ can be Higgsed down to a theory with the
exceptional $G_2$ gauge group and with $N_F'=10$ hypermultiplets in
$({\bf 7})$ living in codimension $56$. Both theories have
$(u,v)=(2,0)$. Let us look for $SO(7)$ and $G_2$ gauge theories with
factorizable anomaly with $(u,v)=(2p,2q)$. In codimension
$64(p+q)-18pq$ one finds an $SO(7)$ theory with $N_A=(p-1)(q-1)$
adjoints $({\bf 21})$, $N_F=3(p+q)-pq$ and $N_S=8(p+q)-4pq$.  In
codimension $56(p+q)-14pq$ one finds a $G_2$ theory with
$N_A'=(p-1)(q-1)$ adjoints and $N_F=10(p+q)-4pq$ fundamentals. The
relations $N_F'=N_F+N_S+N_A-1$ and $N_A=N_A'$ guarantee that these two
theories are connected by the Higgs mechanism for all $(p,q)$.  Again,
following the lines of Section 4, one obtains these models in F-theory
{}from a constrained $D_4$ singularity along a smooth $(p,q)$
curve. Both codimensions turn out to be consistent with such an
interpretation, as does the number of adjoints $N_A$ which is always
given by the genus of the curve.  Again, it is interesting to note
that for $p=1$ the matter spectrum of these mixed
perturbative/non-perturbative theories coincides with that of the
perturbative theory with $12+2q$ instantons.

It is very interesting that, unlike the above example of
$SU(2)_{(m,n)}$, these theories cannot be obtained by Higgsing from
intersecting perturbative and non-perturbative gauge groups. For
instance, there is {\it no} appropriate $SO(7)_{(1,0)}\times
SO(7)_{(0,1)}$ theory with factorizable anomaly form.  In the case of
intersecting $A_n$ type singularities, the fact that they can be
replaced by D-branes implies that we must have a conventional
interpretation of the resulting singularities.  There is no such
reason in the $D_4$ case and evidently we are finding that there must
be new physics going on when such singularities intersect. In fact at
the intersection of the $D_4$ singularities there is a vanishing
2-cycle which signals the appearance of a tensionless string (coming
{}from a 3-brane wrapped around the vanishing cycle).  This is similar
to the occurrence of tensionless strings in strong coupling
transitions in heterotic string theory \sw .

\newsec{Coulomb Branch and Duality Chains}

\lref\rAGM{P.S. Aspinwall, B.R. Greene and D.R. Morrison,
``Calabi-Yau Moduli Space, Mirror Manifolds and Spacetime Topology
Change in String Theory,''
Nucl. Phys. {\bf B416} (1994) 414, hep-th/9309097.}

\subsec{Resolving the singularities}

In our earlier discussion of Tate's algorithm, we indicated what
restrictions on the coefficients in the defining equation would lead
to which kinds of singularities in the total space, but we did not
explain how the singularity type is determined or how the
singularities are resolved.  We will now complete those tasks.

We work with coordinates $x$, $y$, $\sigma$ on the total space, and
wish to recast our conditions on the coefficients in the Weierstrass
equation as being conditions which determine which monomials are
allowed to occur in that equation.  In other words, if we write the
equation in the form
$$\sum c_{i,j,k} x^{i+1}y^{j+1}\sigma^{k+1}=0 ,$$
then we are searching for conditions on $(i,j,k)$ which describe which
monomials are allowed.  The first conditions are that $i\geq-1$,
$j\geq-1$ and $k\geq-1$; we search for other conditions of this form.

Phrasing the problem in this way makes contact with the methods of
toric geometry (see \rAGM\ for a review for physicists).  The
conditions which are natural from the toric point of view are
expressed in terms of vectors $v$ with integer entries, with the
condition given by
$$v\cdot(i,j,k)\geq-1.$$
The initial conditions
mentioned above correspond to the coordinate vectors $(1,0,0)$,
$(0,1,0)$ and $(0,0,1)$, and any additional conditions (with
nonnegative entries in $v$)
automatically correspond to blowups from the toric point of view.
We can resolve the singularity if we find enough such vectors.

For example, the condition associated to $v=(1,1,1)$ can be written as
$i+j+k\geq-1$ and it implies that each allowed monomial
$x^{i+1}y^{j+1}\sigma^{k+1}$ has degree at least two.  In other words,
this is precisely the condition for a singularity to appear at the
origin.  To see the connection to the corresponding blowup,
rewrite a monomial in the form
$$x^{i'}y^{j'}\sigma^{k'}=({x\over\sigma})^{i'}
({y\over\sigma})^{j'}\sigma^{i'+j'+k'}$$
(with the form of this determined by the vectors $(1,0,0)$, $(0,1,0)$,
$(1,1,1)$ used to measure the exponents), and then introduce
$(x/\sigma, y/\sigma, \sigma)$ as coordinates on the blowup.

When applying a condition given by a vector  
$v=(\alpha,\beta,\gamma)$ to a
Weierstrass equation, it should be applied separately to the terms in
the equation, treating each $a_j$ as a polynomial in $\sigma$.  The
corresponding condition will always take the form `$\sigma^k$ divides
$a_j$'. For example, the term $a_3(\sigma)y$ has monomials of the form
$x^{-1+1}y^{0+1}\sigma^{c+1}$ and the condition would imply
$-\alpha+c\gamma\geq-1$ which gives a minimum divisibility for
$a_3(\sigma)$.

We will now work our way through Tate's algorithm, exhibiting the
conditions in toric terms (to the greatest extent possible), and
describing the corresponding blowups.  The first branch of the
algorithm we consider follows the sequence I$_2$, I$_3^{ns}$,
I$_4^{ns}$, \dots.  At the $n^{th}$ step, the condition in Tate's
algorithm is that $\sigma^{[n+1/2]}$ divides $a_3$ and $a_4$, and
$\sigma^n$ divides $a_6$.  It is easy to see that this condition
is reproduced by the vector\foot{We use odd subscripts for
compatibility with a later branch in the algorithm.}
$v_{2k-1}=(k,k,1)$ when $n=2k$, and that it
cannot be expressed in toric terms when $n=2k+1$.

If we perform the
blowups corresponding to $v_1$, $v_3$, \dots, $v_{2k-1}$ in turn,  
we arrive
at a coordinate chart involving $x_k=x/\sigma^k$, $y_k=y/\sigma^k$ and
$\sigma$, and the Weierstrass equation has become
\eqn\blowupeq{
y_k^2+a_1x_ky_k+a_{3,k}y_k=x_k^3\sigma^k+a_2x_k^2+a_{4,k}x_k+a_{6,2k}.
}
The exceptional divisor of the most recent blowup is described by
$\sigma=0$ within this coordinate chart; for generic coefficients, it
is an irreducible nonsingular quadratic equation.  If one additional
power of $\sigma$ divides each of $a_{3,k}$, $a_{4,k}$ and $a_{6,2k}$,
then the exceptional divisor consists of two lines (which will
generally experience monodromy as the coefficients are varied); if in
addition $\sigma^2$ divides $a_{6,2k}$ then there is a singular point
at the origin, leading to the next blowup $v_{2k+1}$.  Thus, in the
generic case we will have found $2k-2$ non-split exceptional divisors
{}from the first $k-1$ blowups, and a $k^{th}$ (split) divisor from
$v_{2k-1}$, giving
the case $A_{2k-1}^{ns}$ (with predicted gauge group $Sp(k)$).  When
there is an additional power of $\sigma$ dividing those three
coefficients, the last step also has $2$ non-split exceptional
divisors and we find the singularity type $A_{2k}^{ns}$ (predicting
unconventional gauge symmetry).  Finally, when $\sigma^2$ divides
$a_{6,2k}$, we must iterate the algorithm again.

The next branch of the algorithm we will follow is I$_2$, I$_3^s$,
I$_4^s$, \dots.  The condition in Tate's algorithm for I$_{2k+1}^s$
can be given torically by the vector $v_{2k}=(k,k+1,1)$.  (Note that
it is possible to give a uniform description of the vectors we have
used so far as $v_n=([{n+1\over2}],[{n+2\over2}],1)$.)  After blowing
up $v_1$, $v_2$, \dots, $v_{2k-1}$, the relevant coordinate chart is
again given by $(x_k,y_k,\sigma)$, with equation again given by
\blowupeq. The condition for I$_{2k+1}^s$ implies that there is one
additional power of $\sigma$ dividing each of $a_2$, $a_4$ and $a_6$,
so the exceptional divisor is described by
$$y_k^2+a_1x_ky_k=0.$$
The vector $v_{2k}$ now gives a toric blowup which in more
conventional terms could be described as blowing up the locus
$\{y_k=\sigma=0\}$. One of the relevant coordinate charts is given by
$(x_k,y_{k+1},\sigma)$, in which the Weierstrass equation becomes
$$y_{k+1}^2\sigma+a_1x_ky_{k+1}+a_{3,k}y_{k+1}=x_k^3\sigma^{k-1}+
a_{2,1}x_k^2+a_{4,k+1}x_k+a_{6,2k+1},$$
with the exceptional divisor for this blowup given by $\sigma=0$ in
this coordinate chart.
If the singularities are no worse than this, the blowups terminate
with an irreducible exceptional divisor.  If they are worse, then
following this branch of the algorithm we have that $\sigma$ divides
$a_{3,k}$ and $a_{6,2k+1}$, and we should blow up $\{x_k=\sigma=0\}$.
Doing so leads back to \blowupeq\ (with the next value of $k$), and
the algorithm then repeats.  It is easy to see that the total process
has produced, in the case of I$_n^{s}$, precisely $n-1$ split
exceptional divisors, so that the singularity type is $A_{n-1}^s$ and
the predicted gauge group is $SU(n)$.

In order to describe the remaining branches of the algorithm
efficiently, we will introduce a bit more toric language.  Toric
geometry teaches us that the combinatorics of a toric resolution of
singularities are essentially determined by the convex hull of the
vectors $v$ used to define the allowed monomials.  Moreover, any integer
vectors which lies in the interior of a codimension one face of that
convex hull represents a toric divisor which does {\it not}\/ meet the
hypersurface defined by the vanishing of the generic allowed
polynomial.  Thus, we can describe the divisors in a toric blowup of
the hypersurface by specifying the vectors $v$ which determine the
convex hull (i.e. the vertices of the convex hull),
as well as all other integer vectors which do not lie in
the interior of codimension one faces (these will lie on {\it edges}\/
of the convex hull).  There are no points in the interior of the
convex hull itself, since we are only resolving singularities which do
not disturb the triviality of the canonical bundle of the space.

We introduce the vectors
$w_k=(k+1,k+2,2)$, and some additional vectors
$u_1=(4,6,3)$, $u_2=(3,5,2)$, $u_3=(6,9,4)$, $u_4=(5,7,3)$.
Then the toric description of these resolutions can be summarized by
the data in the following table.

\vfill
\eject

Table 4:  Toric Data
\bigskip
\begintable
Type | Gauge Group | Vertices | Edge-Vectors \elt
$I_0^{*\,ns}$ | $G_2$ | $w_1$ | $v_2$ \elt
$I_0^{*\,ss}$ | $SO(7)$ | $v_3$, $w_1$ | $v_2$ \elt
$I_{k-1}^{*\,ns}$ | $SO(2k+5)$ | $w_1$, $w_k$ | $v_2$, $v_{k+1}$,
$w_2$, \dots, $w_{k-1}$ \elt
$I_{k-1}^{*\,s}$ | $SO(2k+6)$ | $v_{k+2}$, $w_1$, $w_k$ |
$v_2$, $v_{k+1}$,
$w_2$, \dots, $w_{k-1}$ \elt
$IV^{*\,ns}$ | $F_4$ | $u_1$ | $v_3$, $w_1$, $w_2$ \elt
$IV^{*\,s}$ | $E_6$ | $u_1$, $u_2$, $v_4$ | $v_3$, $w_1$, $w_2$ \elt
$III^*$ | $E_7$ | $w_3$, $u_3$ | $v_4$, $w_1$, $u_1$, $u_2$, $u_4$
\endtable
\bigskip

Note that we have included data for the case of $SO(4k)$, $k>2$,  
even though
we were not able to give a toric description of the conditions for
$SO(4k)$ gauge symmetry.  In fact, by imposing the further divisibilty
condition `$\sigma^{2k+2}$ divides $a_6$', we can force the
factorization
of the corresponding polynomial in Tate's algorithm.  This does not
give a {\it general}\/ polynomial with the corresponding gauge
symmetry, but it does give {\it some}\/ polynomials with that gauge
group.  (This method does not work for $SO(8)$ -- the corresponding
condition would give $SO(7)$ instead). 

There are several additional cases in Tate's algorithm which are not
included in this toric analysis -- cases II, III, IV, and (as already
indicated) case I$_{2k+1}^{ns}$.  In each of these cases, even if the
corresponding condition on the monomials can be described by vectors
$v_j$ those vectors will not have {\it integer}\/ entries.

\subsec{Coulomb Branch and Duality Chains}

We now wish to apply Tate's algorithm and the resolution of
singularities to determine chains of F-theory models (and the
corresponding chains of Calabi--Yau manifolds), related by extremal
transitions, with the transitions mapping to the Higgs mechanism on
the heterotic side.  The starting point is the F-theory model given
in \mv\
which describes the dual of the $E_8\times E_8$ heterotic string with
the instantons distributed as $(12+n,12-n)$.  We describe this model
in terms of the corresponding Weierstrass equation with coefficients
$f$ and $g$ given by \expan.  For our present purposes, we need to
allow the more general form of the Weierstrass equation, which we can
write as
$$\sum c_{ijk\ell}x^{i+1}y^{j+1}z_1^{k+1}z_2^{\ell+1}=0.$$
The conditions which make this a Weierstrass equation can be written
as
\eqn\condi{ 2(i+1)+3(j+1)\leq6,}
while the conditions which restrict the degrees of the polynomials in
$z_1$ and $z_2$ can be written as
\eqn\condii{\eqalign{
k+1&\leq12-4(i+1)-6(j+1)\cr
\ell+1&\leq(12-4(i+1)-6(j+1))+n(6-2(i+1)-3(j+1)-(k+1)) .\cr
}}
These three conditions can be recast in the vector form of the
previous subsection, yielding the vectors
$$\eqalign{
e_5=&(-2,-3,0,0)\cr
e_6=&(-4,-6,-1,0)\cr
e_7=&(-2n-4,-3n-6,-n,-1)\cr
}$$
such that the corresponding conditions take the form
$v\cdot(i,j,k,\ell)\geq-1$.  The standard coordinate vectors $e_1$,
$e_2$, $e_3$, $e_4$ in $R^4$
should be adjoined to these conditions.  (They guarantee that the
exponents in all the monomials are all nonnegative.)

The set of all monomials $x^{i+1}y^{j+1}z_1^{k+1}z_2^{\ell+1}$ which
satisfy the conditions
$$e_\alpha\cdot(i,j,k,\ell)\geq-1,\qquad
\alpha=1,\dots,7$$
forms a so-called {\it reflexive polyhedron}\/ \baty, which is the  
condition
needed to ensure that the generic hypersurface of this type is  
Calabi--Yau.
This was checked explicitly by Candelas and Font \canf, who found the
additional vectors $e_\alpha$, $\alpha>7$, which must be adjoined to
the defining ones in order to completely describe the dual polyhedron
of the polyhedron of monomials.  We will not reproduce those points
here, but we note that for comparison with \canf, one must use the
following change of basis:
$$\eqalign{
e_1&\leftrightarrow(0,0,-1,0)\cr
e_2&\leftrightarrow(0,0,0,-1)\cr
e_3&\leftrightarrow(0,-1,2,3)\cr
e_4&\leftrightarrow(-1,0,2,3)\cr
}$$
(The vectors on the right side are in the notation of \canf.)  Note
that varying the value of $n$ varies the ``top'' of the reflexive
polyhedron in the terminology of \canf.

If the coefficients in the polynomial are generic, then the gauge
group is the one determined in \mv, associated to the $E_8$ factor
with $12-n$ instantons.  For special values of the coefficients,
however, there will be additional singularities of the Calabi--Yau
space along the curve $z_1=0$, which will correspond to gauge symmetry
enhancement in the other $E_8$ factor.  The corresponding polynomials
describe the Coulomb branch for such a gauge group, related to the
original one by an extremal transition.

Calculating these Coulomb branches is a fairly simple matter given all
of the technology we have developed.  For each choice of group in our
first chain, we can give a toric description of the corresponding
moduli space, other than the $SO(8)$ case.
  (In the $SO(12)$ case, the toric moduli space is only a
subspace of the full moduli, but non-toric deformations can be
expected to make up the difference.)  This is done by adding to the
reflexive polyhedron spanned by $\{e_\alpha\}$ certain vectors from
the $u$'s, $v$'s and $w$'s determined in the previous subsection.
(We are implicitly adding a fourth component of $0$ to each of those
vectors, e.g., $v_n$ now denotes $([{n+1\over2}],[{n+2\over2}],1,0)$.)
The results are summarized in table 5.

\vfill
\eject

Table 5:  Chains of Type IIA Duals
\bigskip
\begintable
$H$ | Points to Add | $a_H$ | $b_H$ \elt
$SU(2)$ | $v_1$ | 32 | 24 \elt
$SU(3)$ | $v_1$, $v_2$ | 54 | 36 \elt
$G_2$ | $v_2$, $w_1$ | 54 | 36 \elt
$SU(4)$ | $v_1$, $v_2$, $v_3$ | 76 | 44 \elt
$SO(7)$ | $v_2$, $v_3$, $w_1$ | 76 | 44 \elt
$SU(5)$ | $v_1$, $v_2$, $v_3$, $v_4$ | 100 | 50 \elt
$SO(10)$ | $v_2$, $v_3$, $v_4$, $w_1$, $w_2$ | 124 | 52 \elt
$SO(11)$ | $v_2$, $v_4$, $w_1$, $w_2$, $w_3$ | 124 | 52 \elt
$E_6$ | $v_3$, $v_4$, $w_1$, $w_2$, $u_1$, $u_2$ | 162 | 54 \elt
$E_7$ | $v_4$, $w_1$, $w_3$, $u_1$, $u_2$, $u_3$, $u_4$ | 224 | 56 \elt
$Sp(2)$ | $v_1$, $v_3$ | 64 | 40 \elt
$SO(9)$ | $v_2$, $v_3$, $w_1$, $w_2$ | 96 | 48 \elt
$F_4$ | $v_3$, $w_1$, $w_2$, $u_1$ | 96 | 48
\endtable
\bigskip

(The data in the table corresponds to the ``bottoms'' of the
reflexive polyhedra in \canf; the quantities $a_H$, $b_H$,
defined in \canf, are included to
facilitate easy comparison of results. Note also that ``$pt_1'$'' of
\canf\ is already included in our reflexive polyhedra, as $e_3$, so we
have not included it in the ``points to add'')
This agrees with \canf\ in every particular,
other than in the identification of the gauge groups for some of these
spaces.\foot{Candelas and Font attempted to assign a simply-laced
gauge group to each branch of the moduli space.  Remarkably, there is
always a choice of such a group which produces the correct dimension
of the moduli space.  However, the methods of this paper indicate that
the actual gauge groups are non-simply-laced in several instances.}

\newsec{Conclusion}
We have seen how the enhanced gauge symmetry loci in
the complex moduli space of F-theory (type IIA) compactifications
on Calabi-Yau threefolds get mapped to
the enhanced perturbative, non-perturbative and mixed
 gauge symmetries of the hypermultiplet
moduli for heterotic compactifications on $K3$ ($K3\times T^2$).
On the F-theory side Tate's algorithm proved very helpful
in identifying enhanced gauge symmetry loci.
Using this detailed map we have identified
the Calabi-Yau threefolds dual
to the various possible Coulomb branches of
heterotic compactification on $K3\times T^2$.
This provides us with a systematic method
for mapping out the web of type IIA/heterotic dualities in $d=4$,
$N=2$ theories.

As far as the matter representations are concerned,  more work
needs to be done to verify the structure of the matter
which follows from the duality.  In principle this
should be possible to study: Similar
cases have been analyzed recently in \klkam\
by studying the D-branes of type IIA on Calabi-Yau manifolds.
In this paper, as far as the matter representations go,
we have limited ourselves to what matter
a given singularity must encode in order to be compatible
with duality.  In many cases (especially the simply laced cases)
we have found evidence that the matter is localized
at the zeroes of certain polynomials, extending some of the
observations in \mv.

One can ask whether one can map the F-theory moduli
to the heterotic moduli in a more detailed fashion, even away
{}from the enhanced gauge symmetry points.
In particular it is natural to wonder how the polynomial degrees of  
freedom
we have found on the F-theory side map to the
moduli of bundles on the heterotic side.  Progress
in this direction, as well as a heterotic explanation
of the localization of matter at the zeroes of the polynomials
that we have found, has been recently made \ref\ewit{E. Witten, to  
appear.}.
One can also ask whether we can map the $N=2$, $d=4$ Coulomb phase
on the heterotic side in a more detailed way to the Coulomb
branch on the type IIA side.  Given that we have
identified the relevant Calabi-Yau manifold, one would simply
have to study the Kahler moduli space of this manifold.
However, it is convenient to use mirror symmetry to find the
relevant Calabi-Yau in the type IIB setup:
This can be
done easily as our Calabi-Yau manifolds are nicely characterized
by toric data and thus Batyrev's construction easily applies
to identify the mirror \baty.  We would thus study
the complex structure of this mirror Calabi-Yau and identify it with
the Coulomb branch of the heterotic side.  In particular it should  
be possible
to go to the weak coupling limit of the heterotic string, as in \ketal
\klmvw, and find the field theory
analogues.  Many of these results will be {\it new} even as far as
field theory is concerned, for example $F_4$ with matter.  Note that
given the dictionary we have developed in identifying the various
Coulomb branches, and given mirror symmetry, we have thus managed to
{\it derive} the dictionary for a large number of cases for the
Coulomb branch.  In fact, it is quite suggestive that in the
description of $N=2$ Yang-Mills theories with non-simply laced gauge
groups
\ref\marw{E. Martinec and N.P. Warner, ``Integrable Systems and
Supersymmetric Gauge Theory,'' Nucl. Phys.  {\bf B459} (1996) 97,
hepth/9509161.}, groups appear with a correspondence which is very
similar to what we have found with the outer automorphisms
\outera\ of Dynkin diagrams.  In \marw\ this corresponds to exchanging
the long and short roots.

\vfill
\eject

\centerline{\bf{Acknowledgements}}

We would like to thank P. Aspinwall, J. Lepowski, N. Seiberg, and
E. Witten for valuable discussions.  DRM gratefully acknowledges the
hospitality of the high energy theory group at Rutgers during the
preparation of this work.  The research of MB is supported in part by
NSF grant PHY-92-18167, an NSF 1994 NYI award, and a DOE 1994 OJI
award.  The research of KI is supported in part by NSF grant
PHY-9513835 and the W.M. Keck Foundation.  The research of SK is
supported in part by the Harvard Society of Fellows.  The research of
DRM is supported in part by NSF grant DMS-9401447.  The research of VS
is supported in part by NSF grants DMS 9304580 and PHY 92-45317.  The
research of CV is supported in part by NSF grant PHY-92-18167.

\listrefs
\end